\documentclass{aastex631}

\usepackage{amsmath}
\usepackage{gensymb}
\usepackage{makecell}

\received{January 7, 2023}
\revised{June 1, 2023}
\accepted{5 June, 2023}
\submitjournal{Icarus}

\shorttitle{Evolution of Neptune}
\shortauthors{Chavez et al.}
\graphicspath{{./}{figures/}}

\begin{document}

\title{Evolution of Neptune at Near-Infrared Wavelengths from 1994 through 2022}

\author{Erandi Chavez}
\affiliation{Department of Astronomy, 501 Campbell Hall, 
University of California, Berkeley, CA 94720, USA}

\author{Imke de Pater}
\affiliation{Department of Astronomy, 501 Campbell Hall, 
University of California, Berkeley, CA 94720, USA}
\affiliation{Department of Earth and Planetary Science, 307 McCone Hall, University of California, Berkeley, CA 94720, USA}

\author{Erin Redwing}
\affiliation{Department of Earth and Planetary Science, 307 McCone Hall, University of California, Berkeley, CA 94720, USA}

\author{Edward M. Molter}
\affiliation{Department of Earth and Planetary Science, 307 McCone Hall, University of California, Berkeley, CA 94720, USA}

\author{Michael T. Roman}
\affiliation{School of Physics and Astronomy University of Leicester University Road, Leicester, LE1 7RH, UK}

\author{Andrea Zorzi}
\affiliation{Department of Geological Sciences, Stanford University, Stanford, CA, USA,}

\author{Carlos Alvarez}
\affiliation{W. M. Keck Observatory, 65-1120 Mamalahoa Hwy, Kamuela HI 96743, USA}

\author{Randy Campbell}
\affiliation{W. M. Keck Observatory, 65-1120 Mamalahoa Hwy, Kamuela HI 96743, USA}

\author{Katherine de Kleer}
\affiliation{Division of Geological and Planetary Sciences, California Institute of Technology, Pasadena, CA 91125, USA}

\author{Ricardo Hueso}
\affiliation{Departamento Física Aplicada I, Escuela Ingeniería de Bilbao, Universidad del País Vasco
UPV/EHU, Spain}

\author{Michael H. Wong}
\affiliation{Department of Astronomy, 501 Campbell Hall, 
University of California, Berkeley, CA 94720, USA}

\author{Elinor Gates}
\affiliation{Lick Observatory, P.O. Box 85, Mount Hamilton, CA 95140, USA}

\author{Paul David Lynam}
\affiliation{Lick Observatory, P.O. Box 85, Mount Hamilton, CA 95140, USA}

\author{Ashley G. Davies}
\affiliation{Jet Propulsion Laboratory-California Institute of Technology, Pasadena, CA 91109, USA}

\author{Joel Aycock}
\affiliation{W. M. Keck Observatory, 65-1120 Mamalahoa Hwy, Kamuela HI 96743, USA}

\author{Jason Mcilroy}
\affiliation{The Stratospheric Observatory for Infrared Astronomy (SOFIA), NASA Armstrong Flight Research Center's (AFRC), Palmdale, CA}

\author{John Pelletier}
\affiliation{W. M. Keck Observatory, 65-1120 Mamalahoa Hwy, Kamuela HI 96743, USA}

\author{Anthony Ridenour}
\affiliation{W. M. Keck Observatory, 65-1120 Mamalahoa Hwy, Kamuela HI 96743, USA}

\author{Terry Stickel}
\affiliation{W. M. Keck Observatory, 65-1120 Mamalahoa Hwy, Kamuela HI 96743, USA}

\begin{abstract}

Using archival near-infrared observations from the Keck and Lick Observatories and the Hubble Space Telescope, we document the evolution of Neptune's cloud activity from 1994 to 2022. We calculate the fraction of Neptune's disk that contained clouds, as well as the average brightness of both cloud features and cloud-free background over the planet's disk. We observe cloud activity and brightness maxima during 2002 and 2015, and minima during 2007 and 2020, the latter of which is particularly deep. Neptune's lack of cloud activity in 2020 is characterized by a near-total loss of clouds at mid-latitudes and continued activity at the South Pole. We find that the periodic variations in Neptune's disk-averaged brightness in the near-infrared H (1.6 $\mu$m), K (2.1 $\mu$m), FWCH4P15 (893 nm), F953N (955 nm), FWCH4P15 (965 nm), and F845M (845 nm) bands are dominated by discrete cloud activity, rather than changes in the background haze. The clear positive correlation we find between cloud activity and Solar Lyman-Alpha (121.56 nm) irradiance lends support to the theory that the periodicity in Neptune's cloud activity results from photochemical cloud/haze production triggered by Solar ultraviolet emissions.

\end{abstract}


\section{Introduction} \label{sec:intro}

Neptune is the most distant planet in the solar system, an ice-giant that boasts an active and chaotic atmosphere. The long-term evolution of its atmosphere was first explored with ground-based observations. \citet{lockwood2019} presented a compilation of yearly photometric disk-integrated observations at visible wavelengths (b and y filters) taken from Lowell Observatory between 1972 and 2016 when Neptune was close to opposition. The first set of observations was published in 1986; at the time, the authors reported an anti-correlation in brightness with the solar activity cycle, which they speculated may have resulted from a darkening of aerosols through solar UV irradiation \citep{lockwoodandthompson1986}. In contrast, \cite{moses1989} attempted to explain the perceived correlation by a reduction in the aerosol opacity resulting from variations in the high-energy cosmic rays and consequent ion-induced nucleation.

The planet continued to be monitored on a yearly basis at Lowell Observatory. \cite{sromovsky1993} suggested that the long-term variability in brightness was a seasonal effect, delayed in time by $\sim$30 years. Using the entire database from 1950 to 2005, \cite{lockwoodandjerzykiewicz2006,lockwoodandthompson2002} showed that the observed long-term variability in Neptune's brightness could not be caused by seasonal variations, since the earlier 1950 - 1966 data were much fainter than expected based on seasonal variations. They also reported that the apparent anti-correlation with the solar cycle had "faded", i.e., was no longer present in their more recent data.

The first high-resolution images of the planet were taken during Voyager 2’s 6-month period prior to closest approach in 1989 \citep{smith1989}. These images revealed a dynamic atmosphere characterized by a Great Dark Spot at a planetocentric latitude of $\sim$18$^\circ$S (the GDS), a smaller Dark Spot near 53$^\circ$S (referred to as DS2), a fast moving bright cloud feature ("Scooter") near 41$^\circ$S, and features in the south referred to as South Polar Features (SPF) near a latitude of 70$^\circ$S \citep{smith1989,limayeandsromovsky1991}. 

After Voyager 2, technological advancements led to the Hubble Space Telescope’s (HST) launch in 1990 and the development of Adaptive Optics (AO) at infrared wavelengths for ground-based telescopes in the 1990s. HST and AO were solutions to the atmospheric distortion present in visible and near-infrared data taken by ground-based telescopes using conventional observing techniques. 
Neptune’s appearance in early near-IR observations from the late 1990s and early 2000s was characterized by large and bright mid-latitude bands of activity and a dark equator free of cloud features --- a dramatic change from its appearance in Voyager 2 data \citep[e.g., ][]{max2003,martin2012,roddier1997,gibbard2003,hammelandlockwood1997,sromovsky2001c}. 

\citet{hammelandlockwood2007} gave an excellent summary of Neptune's disk-averaged and disk-resolved observations. They suggested that Neptune undergoes a three-stage brightness pattern in the near-IR, starting with an anomalously bright feature, or "storm", followed by a 5-year period during which a single bright feature dominates the brightness, and ending with a period they refer to as "transitional", where no single feature dominates the brightness. This pattern would explain the higher-than-expected brightness in the mid-70s, and the strong rotational modulation in the near-IR between 1977 and 1980 as reported by \citet{cruikshank1978}, \citet{brown1981}, and \citet{belton1981}. This rotational modulation was absent during 1981-1985, after which time a single feature dominated the brightness again \citep[e.g., ][]{hammel1989solo,hammel1989}. With the Voyager flyby, this feature was recognized as a companion cloud to the GDS \citep[e.g., ][]{smith1989}. Based upon the available data at the time, \citet{hammelandlockwood2007} suggest these years were followed by a quiescent phase, until an anomalously bright feature was recognized in 1993 in Lockwood's photometric data, identified as a bright companion cloud to a new dark spot in early HST data \citep{hammel1995}. Such bright storms were visible throughout the 1990s, observed in high-spatial resolution imaging from the ground through speckle imaging \citep{gibbard2002}.

When AO came online on the 10-m Keck telescope during 1999, both Keck AO and HST images showed a change in the latitudinal distribution of clouds: the clouds were still confined to specific latitudes, but the range of latitudes had broadened considerably, both at mid-southern and northern latitudes, with occasional small features near the equator \citep[e.g., ][]{max2003,karkoschka2011clouds,martin2012}. Moreover, no storms were visible in the early 2000s. A combination of Lowell Observatory \citep{lockwood2019} disk-averaged HST data in blue wavelengths (F467M) \citep{zorzi2019} show that Neptune's brightness continued to gradually increase until southern summer solstice (2005) and remained flat until $\sim$2012, after which it started to decrease overall.

A few noteworthy observations have been reported since 2007: In 2015, a bright storm was seen in near-IR Keck AO data, which was identified as a companion cloud to a new dark spot on the planet \citep{hueso2017,wong2018}, quite similar to the companion clouds seen near the Voyager GDS and dark spots in the 1990s. A second new dark spot was detected in the north in 2018 \citep{simon2019}; this spot was exceptionally large and long-lived \citep{wong2022}. We note that no dark spots were detected during 1996—2015 \citep{hsu2019}.

Potential correlations of variations in Neptune's brightness with changing seasons and the solar activity cycle have been explored, but so far no single cause has been identified. While seasonal effects are most likely important for the slow gradual changes, secular variations in brightness must have a different origin. \citet{aplinandharrison2016} suggest that both UV sunlight and galactic cosmic rays likely affect Neptune's brightness at visible wavelengths. \cite{roman2022} noted a potential correlation between the amount of discrete clouds seen in near-infrared observations \citep{karkoschka2011clouds} and the solar Lyman-alpha flux, but this was based on data limited to less than 2 solar cycles. 

At this point in time, we have almost 30 years of data at high spatial resolution which cover almost 3 solar cycles, although still only $\sim$20\% of Neptune's orbit. In this paper, we investigate changes in Neptune's brightness and cloud cover in these images to address the question of its variability over time. In Section \ref{sec:data} we summarize the (mostly archival) observations from HST, Keck, and Lick from 1994 through 2022 used in this paper; Section \ref{sec:reduc_calib} outlines the data reduction and calibration methods we used for these data. We describe our analysis in Section \ref{sec:cloud_analysis}, which involved several different measurements of Neptune’s cloud activity, including the fraction of the disk covered in clouds, the brightness of the disk, the brightness contribution from the clouds, and the typical pressure levels of these clouds. Finally, we explore overall trends in these quantities that characterize cloud activity, and we speculate about possible causes for any observed patterns.

\section{Data} \label{sec:data}

\begin{figure}[h]
    \centering
    \includegraphics[width=0.75\textwidth]{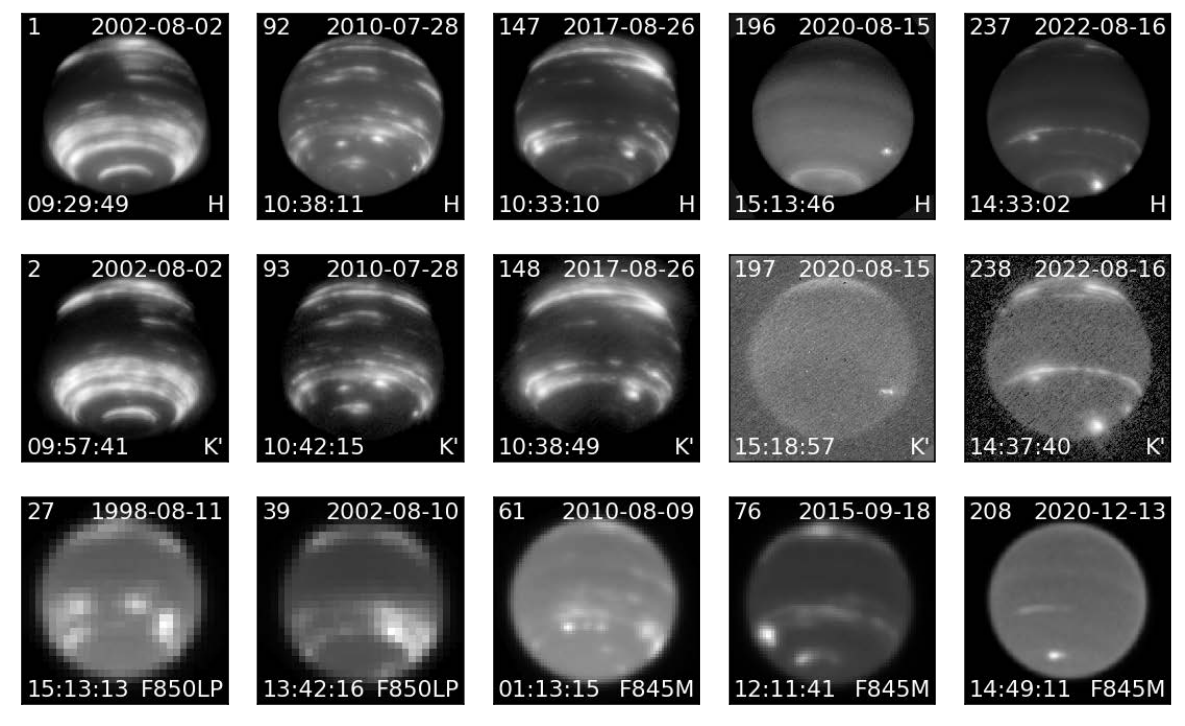}
    \caption{Select example images from Keck (H and K' band) and HST (F850LP and F845M) that display Neptune’s characteristic appearance throughout the three decades worth of data. Note the decreased cloud activity, particularly in the mid-latitudes, after the start of 2020. North roughly points upwards in each image. The index in the top left corner of each image corresponds to the image's index within Figures \ref{fig:hst_pages_of_images} and \ref{fig:keck_pages_of_images} in the appendix, which are collections of the HST and Keck images used in our analysis.
 }
    \label{fig:keck_hst_example_grid}
\end{figure}

Programs that have contributed to long-term observations of Neptune include the Outer Planet Atmospheres Legacy (OPAL)\footnote{\url{https://archive.stsci.edu/prepds/opal/index.html}} program \citep{simon2015} with the Hubble Space Telescope and the Twilight Zone program\footnote{\url{https://www2.keck.hawaii.edu/inst/tda/TwilightZone.html}} \citep{molter2019} at Keck Observatory and Lick Observatory. OPAL data featured in this paper were taken from 2015 through 2021, the Keck Twilight Zone data from 2017 to 2022, and the Lick Twilight Zone data from 2018 and 2019. We included data from these programs in combination with (mostly archival) HST and Keck data to analyze how Neptune’s cloud activity was changing with time. In total, our HST data spanned from 1994 through 2021 and our Keck data spanned from 2002 through 2022, providing nearly 2 decades worth of temporal coverage (see Figure \ref{fig:keck_hst_example_grid} for a few representative images). The total number of observations taken in each filter can be found in Table \ref{tab:filters}.

\subsection{Data Description}

Near-IR data from the Hubble Space Telescope were taken with the Wide Field and Planetary Camera 2 (WFPC2) in FQCH4P15 (893 nm), F953N (953 nm), F850LP (965 nm) and the Wide Field Camera 3 in F845M (845 nm). Data were taken in FQCH4P15 from 1994 to 2002, F850LP from 1996 to 2002, and F953N from 1995 to 2008. A full summary of these filters can be found in Table \ref{tab:filters}. Images with the WFPC2 were sporadic, with observations taking place during 1-10 days scattered throughout each year. Each of these dates had a few observations taken. Following the WFC3's installation in 2009, F845M data were taken between 2009 and 2021. The WFC3 data were taken 1-5 days every year, with multiple images being taken during each date. Much of the data during and after 2015 were taken as part of the OPAL Program \citep{simon2015}. When taking both the WFPC2 and the WFC3 images into account, the data were regularly sampled on a yearly basis from 1994 through 2021, with the exception of 1999, 2003, and a 3-year gap from 2012 to 2014. WFC3 data from 2018, 2019, and 2020 were magnified to pixel scales 2-3 times smaller than the standard 39.62 mas/pixel by \cite{wong2022}; note, though, that this process does not increase the resolution of the images. The 2021 data were not magnified.

Near-infrared images were taken with the Near Infrared Camera 2 (NIRC2) on the Keck Telescopes from 2002 through 2022 in H band (1.63 $\mu m$) and K’ band (2.12 $\mu m$), and the Shane AO infraRed Camera and Spectrograph (ShARCS) on the Shane Telescope at Lick Observatory during 2018 and 2019 in H band (1.66 $\mu m$); see Table \ref{tab:filters} for filter descriptions. These telescopes both utilized adaptive optics (AO) systems. Keck NIRC2 and Lick ShARCS had pixel sizes of 9.94 mas \citep{depater2006} and 33 mas \citep{mcgurk2014lick}, respectively.

Between 2002 and 2017, the Keck data were typically taken on a few consecutive nights each year, producing regularly sampled yearly data. In 2017, the Twilight Zone Program was implemented at Keck Observatory, giving classically scheduled observers the opportunity to donate their unused telescope time (often during poor weather conditions or twilight hours) to take short (10 - 40 minute) observations of bright solar system objects \citep{molter2019}. The addition of Twilight program data increased the temporal frequency of Keck data between 2017 and 2022, with gaps between some observations being shortened to weeks, or days. Since its implementation in 2015, the Twilight Zone Program has also been active at Lick Observatory, resulting in data taken during 2018 and 2019 \citep{molter2019}.

\begin{deluxetable}{cccccccp{19mm}}
\tablecaption{List and description of telescope filters. \label{tab:filters}}
\tablewidth{0pt}
\tablehead{ 
\colhead{Telescope} & \colhead{Filter} & \colhead{Central Wavelength} & \colhead{Range} & \colhead{Pixel Scale} & \colhead{Cutoff Fraction} & \colhead{Image Count} & \colhead{Dates} \\
 & & \colhead{$\lambda$ ($\mu m$)} & \colhead{$\Delta \lambda$ ($\mu m$)} & \colhead{(mas/pixel)} & \colhead{$X_c$} & &
}
\startdata
Keck NIRC2 & H & 1.63 & 0.30 & 9.942 & 1.5 & 149 & 2002-08-22 to 2022-09-12 \\
Keck NIRC2 & K' & 2.12 & 0.35 & 9.942 & 2.5 & 118 & 2002-08-22 to 2022-09-12 \\
Lick ShARCS & H & 1.66 & 0.30 & 33 & 1.5 & 10  & 2018-09-27 to 2019-09-15\\
HST WFPC2 & FQCH4P15 & 0.8930 & 0.0055 & 99.6 & 1.5 & 22 & 1994-06-28 to 2002-08-10 \\
HST WFPC2 & F953N & 0.9546 & 0.00525 & 99.6 & 1.5 & 30 & 1995-09-01 to 2008-07-23 \\
HST WFPC2 & F850LP & 0.9650 & 0.16724 & 99.6 & 1.5 & 6 & 1996-08-13 to 2002-08-10 \\
HST WFC3 & F845M & 0.8454 & 0.0870 & 39.62 & 1.25 & 176 & 2009-10-22 to 2020-12-13 \\
\enddata
\tablecomments{2018, 2019, and 2020 F845M data were magnified to pixel scales 2-3 times smaller than the standard 39.62 mas/pixel by \cite{wong2022}.}
\end{deluxetable}

\subsection{Data Reduction and Calibration \label{sec:reduc_calib}}

All frames of Keck and Lick data underwent standard image reduction that involved sky subtraction, flat fielding, and median value masking to remove bad pixels. A dither pattern with Neptune positioned in different regions of the detector was used to account for detector artifacts and construct a sky background after median averaging the frames. A three-point dither was used for Keck observations and a five-point dither was used for Lick observations. After cropping and aligning the images, the frames were median averaged to produce a final image.

The Keck data were usually reduced with in-house IDL routines, and the geometric distortion of the array was corrected using the dewarp routines provided by Brian Cameron of the California Institute of Technology\footnote{\url{http://www2.keck.hawaii.edu/inst/nirc2/forReDoc/post-observing/dewarp/nirc2dewarp.pro}}. Keck and Lick data taken with the Twilight Zone Project were reduced in Python, using the package nirc2\_reduce \citep{nirc2reduce}.  

When possible, the Keck data were photometrically calibrated using known photometric stars \cite[from e.g.,][]{elias1982,cutri2003vizier,leggett2006,hunt1998} (see also individual papers of previously published images, Table \ref{tab:keck_lick_data_table}). 
All calibrated data were converted to units of I/F, as defined by \cite{hammel1989}:

\begin{equation}
    \frac{I}{F} = \frac{r^2}{\Omega}\frac{F_{N}}{F_{\odot}} \label{eq:if_eq}
\end{equation}

where r is Neptune's heliocentric distance in A.U., $\pi F_{\odot}$ is the Sun's flux density at Earth's orbit (1 A.U.) \citep[from][]{colina1996}, $F_{N}$ is Neptune's observed flux density, and $\Omega$ is the solid angle corresponding to a single pixel on the detector.

HST data were reduced and calibrated in the manner described by \citet{wong2018,wong2022}, and were converted to units of I/F using Equation \ref{eq:if_eq}. The HST data from 2020 were deconvolved using the method described in \citet{fryandsromovsky2023}.

\subsubsection{Calibration Method for Non-Photometric Keck Data \label{sec:bootstrap}}

While some Keck data were photometrically calibrated, the images taken as part of the Twilight Zone program at Keck and Lick were not calibrated; neither were images taken during nights that were not photometric. However, we were able to calibrate the remaining Keck data using a different calibration method that relied on assumptions about how Neptune’s background reflectivity changed with time. 

At near-IR wavelengths, all light contributions are due to reflected sunlight from clouds and hazes, not thermal emission from the planet. We took Neptune’s background to be cloudless regions on the disk where light contribution is due to reflective haze layers. \citet{zorzi2019} showed that Neptune’s mean background brightness in HST methane-band data (F850LP and F845M) between 1996 and 2019 changed over time, however these occurred over the span of multiple years and were minimal in comparison to changes in the disk-averaged I/F \citep[see Figure 4.2 in][]{zorzi2019}. If Neptune's background behaved similarly in H band and K' band, we could assume that Neptune's background in both filters remained relatively constant over the span of 1-2 years. We made this assumption when proceeding with the calibration for these data.

\begin{figure}[h]
    \centering
    \includegraphics[width=0.75\textwidth]{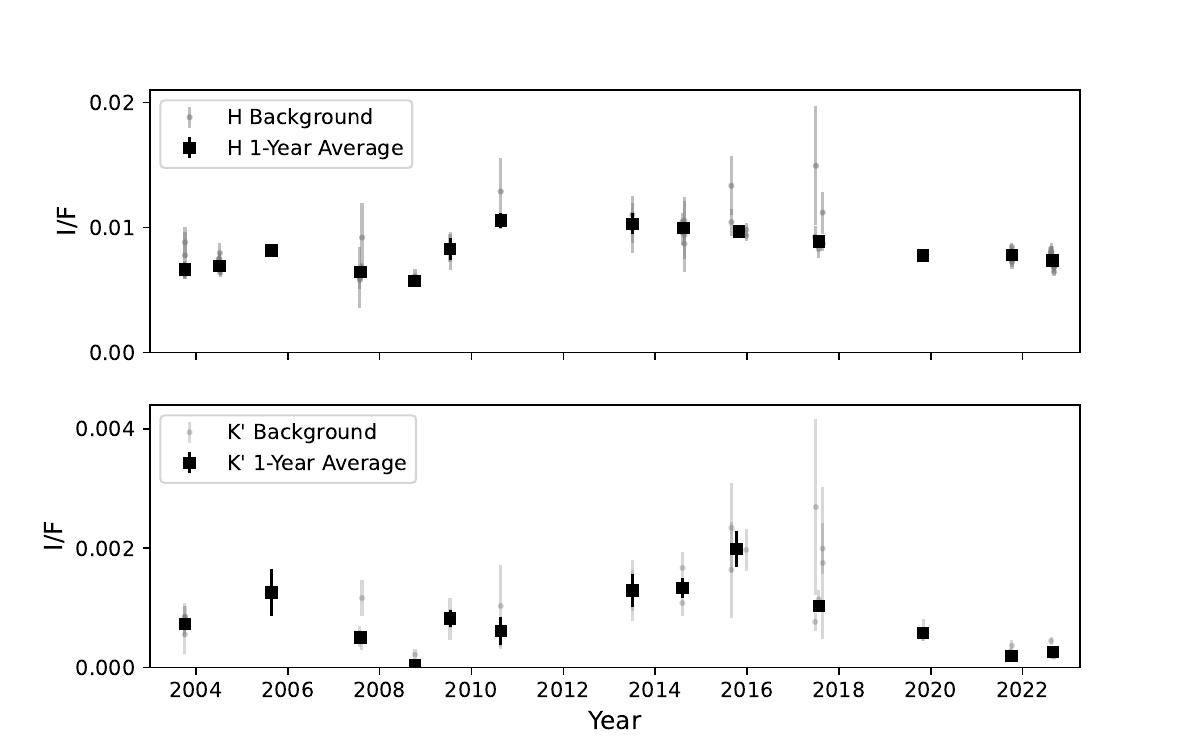}
    \caption{Neptune’s background I/F (light gray points), and the yearly weighted average of the background I/F (black squares) as determined by photometrically calibrated Keck data. The background I/Fs and the associated $1 \sigma$ error were determined by taking the mean and standard deviation of a cloudless equatorial region within an observation (see Section \ref{sec:bootstrap}). The yearly weighted averages were used to calibrate the rest of the Keck data (see Section \ref{sec:bootstrap} for description of calibration method). Top: Results from H band data. Bottom: Results from K’ data. }
    \label{fig:avg_h_kp_background}
\end{figure}

The typical background I/F of Neptune in an image was calculated by taking the median of a square region within the disk that contained no cloud features. We used the equatorial region of Neptune, as it was typically free of cloud features throughout all filters (see Figures \ref{fig:keck_pages_of_images} and \ref{fig:hst_pages_of_images}). The size of the box was chosen to span a fraction of Neptune's equatorial band, spanning several degrees of latitude and longitude on the planet’s surface. The box's location on the disk was consistent throughout all images within a filter to avoid the effects of limb darkening impacting our background calculation (see Figure \ref{fig:bootstrap_box_example} for an example of box placement). The standard deviation of the data within the box was used as the 1-$\sigma$ error on the background I/F. The background I/F for the calibrated H and K' band data is shown in Figure \ref{fig:avg_h_kp_background}.

Figure \ref{fig:avg_h_kp_background} shows that Neptune’s background I/F within a year are typically clustered around similar values. We assumed that Neptune’s background remained constant over the span of a year, and determined the background I/F for each year by taking the weighted average of the background I/F values for each year in photometrically calibrated H and K’ data; the results are shown in Figure \ref{fig:avg_h_kp_background}. We used the error on the weighted mean as the $1 \sigma$ error. A weighted average was used to decrease the impacts that large storms could have on the background I/F measurement. For example, the increased spread during 2017 may have been caused by contamination from the large equatorial storm observed during that year \citep{molter2019}, and we therefore expect that the presence of other large storms could have an impact on determining the background I/F. We then used these yearly-averaged backgrounds to photometrically calibrate the uncalibrated Keck data. From the assumption that Neptune’s background remained constant over the span of 1-2 years, we multiplied the uncalibrated data by the ratio between the nearest yearly-averaged background I/F and its background I/F, which calibrated the image. Since all uncalibrated Keck images had a yearly-averaged calibrated background I/F taken within 2 years, we were able to calibrate the previously uncalibrated Keck data with this method. We did not apply this method to the Lick H band data due to the lack of photometrically calibrated Lick images. While we could have used the backgrounds as determined from Keck H band data to calibrate the Lick data, the differences in the spatial resolution and PSF between the Keck and Lick data would introduce major uncertainties in the resulting calibration. Therefore, the Lick data remained uncalibrated.

\section{Cloud Activity Analysis} \label{sec:cloud_analysis}

As pointed out in the Introduction, while seasonal effects are likely important, the cause of secular variations in Neptune's brightness is not known; both UV sunlight and galactic cosmic rays have been suggested. With almost 30 years of data at high spatial resolution from HST and Keck combined, we use these data to further investigate the cause of Neptune's variability. To do so, we first determine the fraction of Neptune's disk that was covered by clouds, and how this varies over time. Our second method consists of measuring the average brightness or reflectivity (I/F) of Neptune's disk and clouds in an image. We then separate these I/F contributions by clouds from an average background haze, and then ``normalize'' the disk-averaged I/F at the different wavelengths to account for wavelength-dependent variations in reflectivity.

\subsection{Fractional Cloud Coverage \label{sec:frac_cloud_cover}}

We measured the fractional cloud coverage on the planet's disk by determining a ``cutoff'' value for which all regions with a brightness higher than this value were determined to be clouds. This approach did not require calibrated data to determine the cloud coverage. 
We began with Neptune’s background brightness measurements in all images, determined by the equatorial box method outlined in Section \ref{sec:bootstrap}. We then determined a cutoff value, $X_c$, such that all pixels with values higher than $X_c \times$background were determined to be clouds. The precise value of $X_c$ did depend on the filter in question, since the contrast between the clouds and background was different for different filters. For example, a higher $X_c$ was necessary for K’ band than H band, as the contrast between cloud features and the background was much higher in K’ band. The cutoff values $X_c$ used for each filter are listed in Table \ref{tab:filters}. These were chosen to be low enough that major cloud features throughout the disk, including near the limbs were captured, but high enough so that background hazes were not counted as cloud features. This meant that faint features were not captured as effectively as bright, prominent ones. Figure \ref{fig:bootstrap_box_example} shows an example of the cloud features on an image determined with this cutoff method.

\begin{figure}[h]
    \centering
    \includegraphics[width=0.95\textwidth]{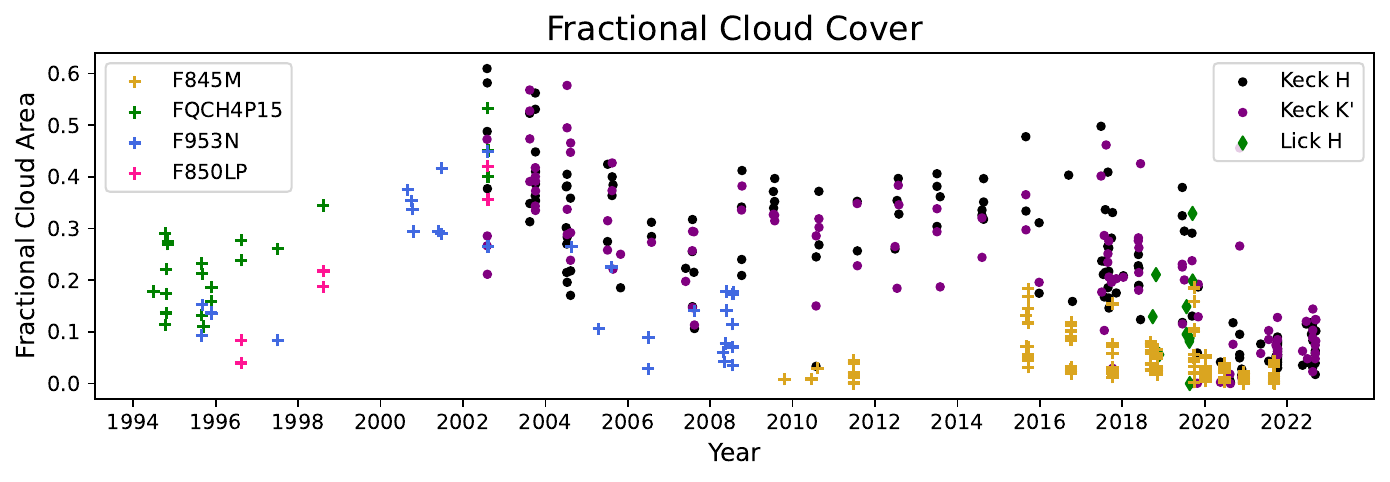}
    \caption{The fractional area of Neptune's disk that contained clouds. This area was determined by counting the number of pixels on Neptune's disk above a ``cutoff value'' and dividing it by the total number of pixels subtended by the disk in an image. The HST filters of F845M, FQCH4P15, F953N, and F850LP are shown with gold, green, blue, and pink crosses respectively. Keck H band and K' band results are shown with black and purple circles, respectively. Lick results are shown with green diamonds. }
    \label{fig:cloud_cover}
\end{figure}

After determining our choices for these cutoff values, we isolated Neptune's disk in every image. We did this by using an ellipse that traced the edge of the planet. The semimajor axis of this ellipse was used as Neptune's equatorial radius ($R_p$), while the semiminor axis was used as Neptune's apparent polar radius: $R_{P,app} = \sqrt{R_{eq}^{2} \sin^2{\phi} + R_{p}^{2} \cos^2{\phi}}$ where $\phi$ was the sub-Earth latitude for the given observation, which was retrieved from JPL Horizons \citep{karim2018}. Once we determined this ellipse, we summed up the number of pixels it contained, which we used as the total pixels spanned by Neptune's disk ($N_{d}$). We then calculated the number of pixels within Neptune’s disk that were above the associated cutoff value $(N_{c})$. The ratio between these two values was used as the fractional cloud coverage $F_c$ for Neptune in the image: $F_{c} = N_{c} / N_{d}$. Our fractional cloud coverage results from all filters are shown in Figure \ref{fig:cloud_cover}.

\subsection{Disk-Averaged I/F \label{sec:disk_avg_if}}

\begin{figure}[h]
    \centering
    \includegraphics[width=0.9\textwidth]{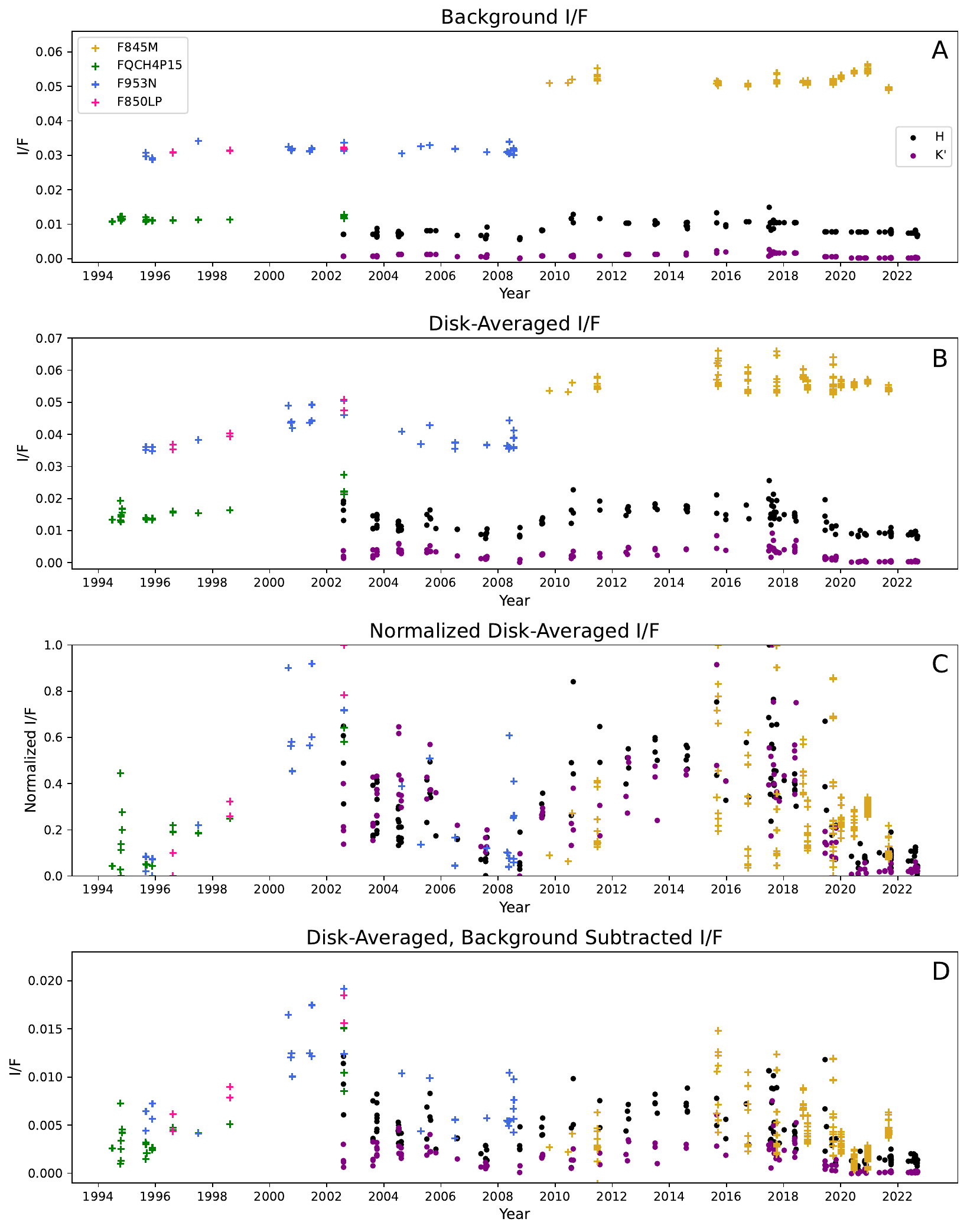}
    \caption{Data shown in all 4 panels include H band (black circles), K’ band (purple circles), F845M (golden crosses), FQCH4P15 (green crosses), F953N (blue crosses), and F850LP (golden crosses) from 1994 to 2022, as available. All H and K’ band data were either photometrically calibrated using standard methods or were calibrated using the method described in Section \ref{sec:bootstrap}. Lick data are not included, as they were not calibrated. A: The background I/F of Neptune measured in all 6 filters. Note that the H and K’ background I/F that were determined from photometrically calibrated Keck data shown here are also shown as light gray points in Figure 2. The rest of the H and K’ background I/F measurements were from data calibrated with the method described in Section \ref{sec:bootstrap}. B: The disk-averaged I/F of Neptune in all 6 filters. C: The disk-averaged I/F of Neptune in panel B, now normalized within each filter for better comparison. D: Neptune's disk-averaged I/F (panel B) after subtraction of the corresponding background I/F measurement (panel A). This isolates the contributions to Neptune’s reflectivity due to clouds.}
    \label{fig:if_process}
\end{figure}

We calculated the disk-averaged I/F of Neptune in our photometrically-calibrated Keck data. We began by isolating the planet from the background sky by drawing a circle centered on Neptune’s disk on each image. The radius of this circle ranged between 10 and 20 pixels larger than Neptune's equatorial radius to account for brightness that was spread past the edge of Neptune’s disk due to its point-spread function (PSF). The radius was increased more for dates with poor weather or when massive, bright features appeared on the disk, both of which caused a large amount of light to be spread past the disk edge. This was a common occurrence in the H and K' band data (see Figures \ref{fig:keck_pages_of_images} and \ref{fig:lick_pages_of_images}). 

Summing up the brightness within the circle gave Neptune's disk-integrated I/F. Converting the disk-integrated I/F to disk-averaged required us to divide by the total number of pixels on Neptune's disk, see Section \ref{sec:frac_cloud_cover} for how we determined this. This process was repeated for every calibrated Keck and HST image in all filters, resulting in Figure \ref{fig:if_process}.

\subsection{Disk-Averaged, Background-Subtracted I/F}\label{sec:Background-SubtractedIF}

As described earlier, the disk-averaged I/F includes contributions from the background aerosol hazes/cloud-free regions, as shown in the images and in Figure \ref{fig:if_process}. Although there are latitudinal differences in the temporal evolution of Neptune's background, these are relatively small in comparison to the changes in the disk-averaged I/F (see Figure \ref{fig:zorzi2019diskvsbackground}) and hence by subtracting the background I/F from the disk-averaged I/F, we can remove the total background contribution to the first order. The results are shown in Figure \ref{fig:if_process}.

While panel D of Figure \ref{fig:if_process} appears closer to a direct comparison of cloud activity across multiple filters due to the removal of background contributions, we made one final correction: - taking the wavelength-dependent cloud spectra. For example, even when the same feature was observed in both filters, the H band consistently showed a higher disk-averaged, background-subtracted I/F than the K' band. This discrepancy between filters could obscure underlying patterns of activity. We, therefore, corrected the background-subtracted, disk-averaged I/F to be spectrally flat; the procedure for doing so is described in Section \ref{sec:flatspectrum}.

\subsection{Modeling Cloud Pressures, Spectrally Flat Disk-averaged I/F \label{sec:flatspectrum}}

Modifying the typical background-subtracted, disk-averaged I/F in panel D of Figure \ref{fig:if_process} to be spectrally flat requires radiative transfer modeling. We use the SUNBEAR (Spectra from Ultraviolet to
Near-infrared with the BErkeley Atmospheric Retrieval) radiative transfer program to model Neptune's atmospheric spectrum \citep[for details, see][]{molter2019}. To understand how a cloud's brightness changes depending on the filter it was observed in, we need to generate a cloud spectrum using SUNBEAR that represents the typical clouds in Figure \ref{fig:if_process}. All modeling is performed at $\mu = 0.67$ to simulate a disk-averaged value, where $\mu$ is the cosine of the emission angle.

We first determine the typical cloud pressures at which observed cloud features were located. Having obtained a substantial amount of data in both the H and K' band, we used the following equation \citep{depater2011}:

\begin{equation}
    \frac{I_{c, K'} - I_{b, K'}}{I_{c, H} - I_{b, H}} = \frac{I_{K'}(P_{m}) - I_{b,K'}}{I_{H}(P_{m}) - I_{b,H}}
    \label{eq:kp_over_h}
\end{equation}

\begin{figure}[h]
    \centering
    \includegraphics[width=0.5\textwidth]{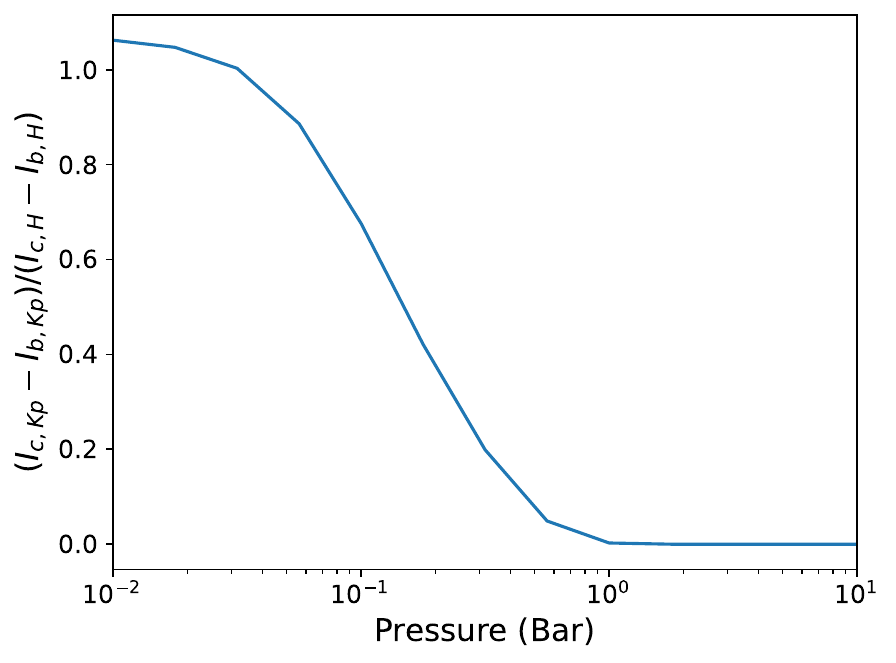}
    \caption{The relationship between cloud pressure and the ratio of that cloud's background-subtracted I/F in K' and H band as determined by Equation \ref{eq:kp_over_h}.}
    \label{fig:kp_h_pressure}
\end{figure}

In this equation, the ratio between the intensity of a background-subtracted cloud ($I_c - I_b$) observed in K' and H band is related to the pressure of that cloud, $P_m$. Assuming an optically thick cloud at different pressure levels in the atmosphere, we used a radiative transfer model to solve Equation \ref{eq:kp_over_h} \citep[for details, see][]{molter2019}; the results are shown in Figure \ref{fig:kp_h_pressure}.

\begin{figure}[h]
    \centering
    \includegraphics[width=0.90\textwidth]{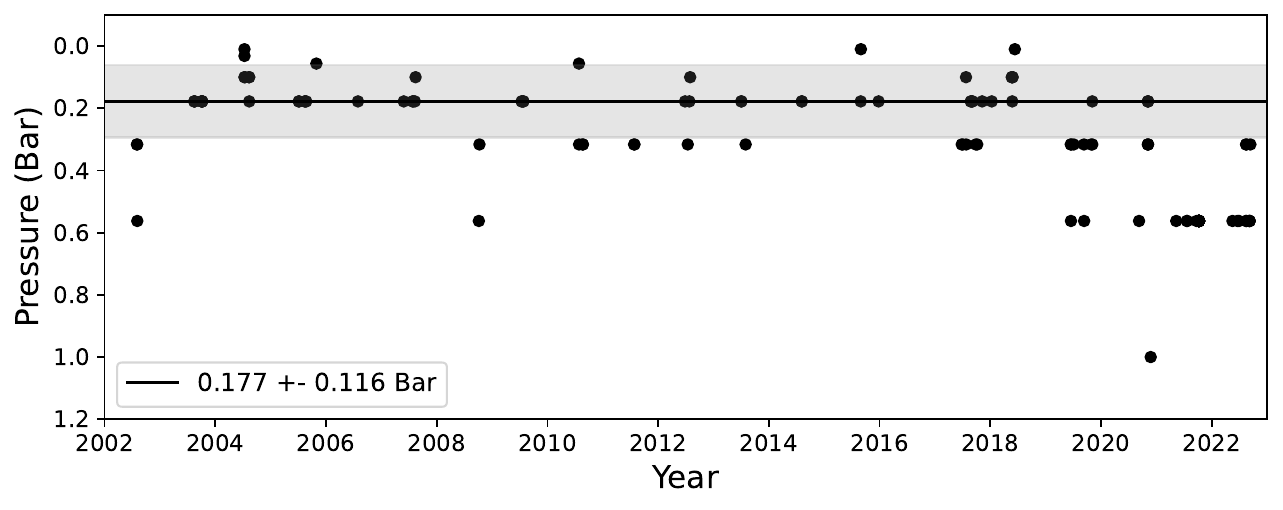}
    \caption{The average cloud pressure of all cloud features that were observed in both Keck H and K' band. This was calculated by taking the background-subtracted, disk-averaged I/F in Keck K’ and H band data, and using the conversion shown in Figure \ref{fig:kp_h_pressure} to convert the ratio of the two to cloud pressure. The median of these pressures from 2002 through 2019, 0.177 Bar, is shown with a black line. The gray shaded region represents the 1-sigma error of 0.166 Bar.}
    \label{fig:cloud_pressure}
\end{figure}

This process required that the same features be captured in both H and K’ band. Neptune hosts many rapidly (timescales $<$1 hr) evolving features \citep{limayeandsromovsky1991,sromovsky1993,martin2012}. To maximize the chances of capturing the same features in both H and K' band images to determine the clouds' pressure levels with Equation \ref{eq:kp_over_h}, we limited our selection to calibrated H and K’ band images taken within 1 hour of each other. This meant that the Lick and HST data were not utilized in this step. Equation \ref{eq:kp_over_h} also required that the cloud feature be visible in H band, so we only used data where the H band image had a non-zero disk-averaged, background-subtracted I/F. After these images were identified, we divided the K’ and H band background-subtracted, disk-averaged I/F, and used Figure \ref{fig:kp_h_pressure} to convert this value to atmospheric pressure. The results shown in Figure \ref{fig:cloud_pressure} thus shows the typical cloud pressure on Neptune at that time. As shown, these pressures vary between $<$0.1 bar down to $\sim$0.3 bar, sometimes down to 0.6 bar and once to 1 bar. This agrees well with the pressures found for individual clouds by \citet{gibbard2003}, \citet{depater2014} (their Table 8), and \citet{irwin2011}. Clouds in the north have typically been found in the stratosphere, at mid-southern latitudes both in the stratosphere and troposphere, and clouds at high southern latitudes (including SPFs) are usually somewhat deeper in the atmosphere.

\begin{figure}[h]
    \centering
    \includegraphics[width=0.95\textwidth]{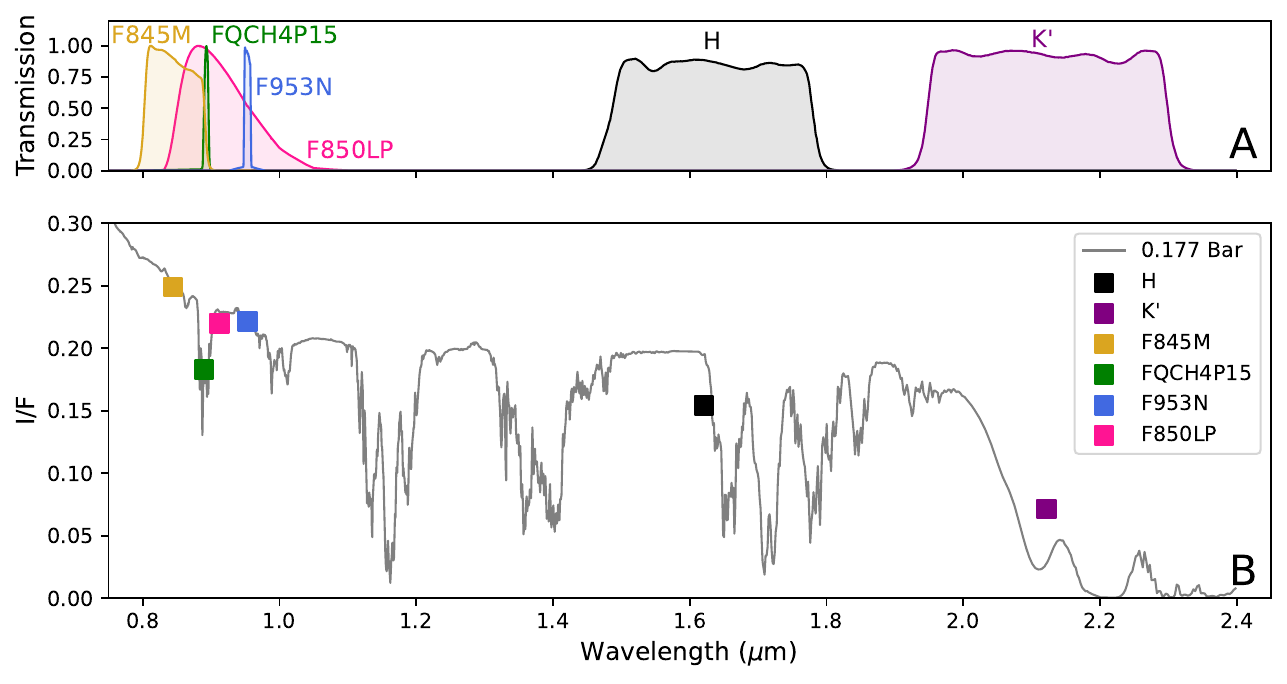}
    \caption{A) Filter transmission curves for HST (F845M, FQCH4P15, F953N, and F850LP) and Keck (H and K') filters. B) The cloud spectra for a cloud at 0.177 bar obtained with the SUNBEAR radiative transfer program (see Section \ref{sec:Background-SubtractedIF}), the pressure at which typical clouds observed in H and K' band appeared between 2002 and 2022. Overplotted in squares are the results from convolving the filter transmission curves in panel A with the cloud spectra. These were used to scale the data in panel D of Figure \ref{fig:if_process} to be spectrally flat.}
    \label{fig:cloud_spectrum_convolved}
\end{figure}

We took the median of the disk-averaged cloud pressure for data from 2002 through 2019 and found it to be 0.177 bar (see Figure \ref{fig:cloud_pressure}); this value was used as input to our spectral flattening procedure outlined below. We did not consider the data from 2020 onwards due to an increase in typical cloud pressures, which is discussed more in Section \ref{sec:results}. The resulting cloud spectrum is shown in Figure \ref{fig:cloud_spectrum_convolved}.

\begin{figure}[h]
    \centering
    \includegraphics[width=0.95\textwidth]{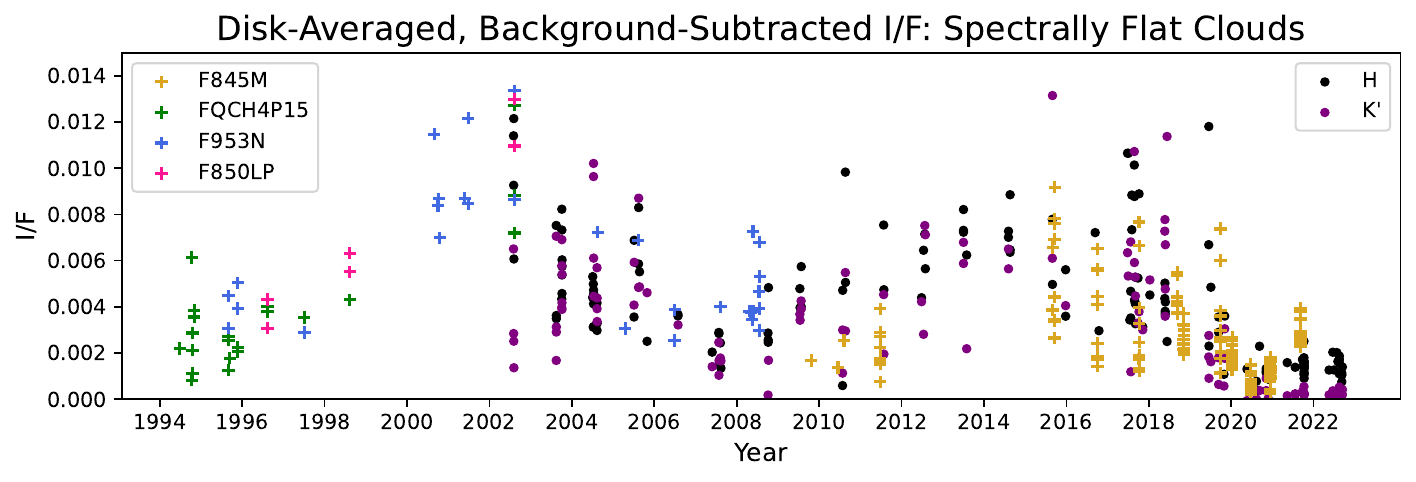}
    \caption{The disk-averaged I/F subtracted by the background I/F of Neptune's disk scaled to be spectrally flat. Each filter's data is scaled to match the H band using the results from convolving each filter's transmission curve with a cloud spectrum at 0.177 bar (see Figure \ref{fig:cloud_spectrum_convolved}).}
    \label{fig:spec_flat_if}
\end{figure}

For each HST and Keck filter in our data base, we convolved the spectrum in Figure \ref{fig:cloud_spectrum_convolved} with the filter transmission curves. Given this modeled spectrum, the colored squares in Figure \ref{fig:cloud_spectrum_convolved} represent the cloud's I/F value we expect to observe in each filter. We used this to scale the background-subtracted, disk-averaged I/F from panel D of Figure \ref{fig:if_process} to make them spectrally flat. We normalized all values to the H band, following a similar approach to \cite{roman2022}. For each filter, we calculated the ratio between the expected I/F value of the H band and the expected I/F value of that filter: $[I/F]_{H} / [I/F]_{f}$. Multiplying each filter's disk-averaged, background subtracted I/F by its associated ratio resulted in the spectrally flat disk-averaged, background subtracted I/F. The results are shown in Figure \ref{fig:spec_flat_if}; we used this quantity as a measure of Neptune’s disk-averaged cloud brightness over time.

\section{Results and Discussion \label{sec:results} }

\subsection{Fractional Cloud Area and Cloud I/F: Overall Patterns}

The two quantities we use to measure cloud activity are the fraction of Neptune's disk that contains clouds (see Figure \ref{fig:cloud_cover}) and the spectrally flat disk-averaged, background-subtracted I/F (see Figure \ref{fig:spec_flat_if}). In this section, we refer to these two quantities as Neptune's cloud coverage and average cloud brightness. 

The results of our cloud coverage work should be compared with the work done by \cite{karkoschka2011clouds}. That author analyzed Neptune's cloud coverage between 1994 and 2010 in HST data, and while they used slightly different filters, including some at visible wavelengths, their results show the same general pattern as ours: a rise and subsequent fall in cloud coverage, with minima in $\sim 1995$ and $\sim 2007$ and a maximum in $\sim 2002$ \citep[see][Figure 9]{karkoschka2011clouds}.  We extend this time record using HST, Keck and Lick data, finding another maximum in cloud coverage in $\sim 2015$ and a deep minimum during 2020-2022.

Averaging the cloud coverage and the average cloud brightness over year-long bins and plotting these quantities against each other (see Figures \ref{fig:binned} and \ref{fig:p_lya_cloud_if}) confirms that these quantities are correlated and show the same periodic pattern. However, we will proceed with the average cloud brightness as the more accurate measurement of cloud activity. Since the determination of cloud coverage may suffer from ``bleeding'' effects of cloud features due to the PSF, a more accurate way to analyze potential variations is by using our average cloud brightness. For example, the discrepancy in cloud coverage in $\sim$2007 between HST and Keck data is not present in the average cloud brightness, which we attribute to PSF contamination. Two distinct peaks in activity occur in $\sim 2002$ and $\sim 2015$, and two minima are seen around 2007 and 2020, suggesting that the cloud I/F has a $\sim 13$ year period variation (see Figure \ref{fig:spec_flat_if}). The first maximum in 2002 appears to be somewhat brighter than the second (2015) peak.  The precise timing of the second maximum, however, is difficult to determine due to the large scatter in the data between 2017 and 2019. There clearly is a dramatic drop, however, in cloud I/F during late 2019/early 2020 in all 3 filters. This mirrors the drop in cloud coverage during the same time period. 

The time sampling of the data increased in 2015 and 2017 due to the OPAL and Twilight Zone programs, respectively. With more data, daily variations in cloud activity are better documented. These variations show high levels of scatter in cloud coverage and average cloud brightness during 2017 and 2019 - years where large storms were seen on Neptune. In particular in 2017 a large, bright equatorial storm was discovered \citep{molter2019}, and in 2019 several long-lived storms were observed at mid-latitudes \citep{chavez2022}.

The comparison between the different panels of Figure \ref{fig:if_process} show that patterns in disk-averaged brightness were primarily driven by cloud activity. Panel C of Figure \ref{fig:if_process} shows that the disk-averaged I/F within each filter displays the same periodic pattern as the average cloud brightness (Figure \ref{fig:spec_flat_if}) and the cloud coverage (Figure \ref{fig:cloud_cover}): maxima during 2002 and 2015, minima during 1995, 2009, and 2020. These patterns are not reflected in Neptune’s Background I/F (panel A of Figure \ref{fig:if_process}), which stayed relatively constant over time. Therefore the changes in Neptune’s disk-averaged brightness during this time were primarily driven by changes in the discrete cloud activity, and not by changes in the background hazes.

\subsection{Drop in Cloud Activity During Late 2019/Early 2020}

During the second peak in cloud activity ($\sim 2013 - 2019$), the mid-latitudes were very active, hosting bands of activity and large, long-lived features that persisted for months \citep[][see Figures \ref{fig:hst_pages_of_images} and \ref{fig:keck_pages_of_images}]{hueso2017,chavez2022}. However, Figures \ref{fig:hst_pages_of_images} and \ref{fig:keck_pages_of_images} reveal a dramatic transition in Neptune's cloud activity that began during late 2019. This transition is characterized by the loss of mid-latitude cloud activity, and by early 2020, cloud activity at mid-latitudes has nearly disappeared. In stark contrast to its appearance in previous years, Neptune's typical appearance during 2020 is dominated by background hazes with an active south pole and occasional small, isolated cloud features at mid-latitudes. 

Images of decreased cloud activity characteristic of 2020 were taken as early as September and October 2019 (see Images 134 and 135 from Figure \ref{fig:hst_pages_of_images} and Image 176 from Figure \ref{fig:keck_pages_of_images}). However, subsequent images during November 2019 still show high levels of cloud activity at southern mid-latitudes (see Images 178-180 from Figure \ref{fig:keck_pages_of_images}), demonstrating a gradual transition in cloud activity. HST images from January 7 and 8, 2020, show far less activity than those from late 2019, with some isolated mid-latitudinal features present. Subsequent H and K' band Keck data from May 2020 and F845M data from June 2020 show an even emptier disk, and represent the typical appearance of Neptune during 2020 - 2022. 

This dramatic shift was captured in our cloud activity analysis through the dip in cloud coverage and in the disk-averaged cloud brightness during 2020 (see Figures \ref{fig:cloud_cover} and \ref{fig:spec_flat_if}), both of which resulted from the low number of cloud features on the disk. The lack of scatter in cloud activity typical of previous years further confirms that Neptune's disk consistently lacked clouds during this period of low activity. 

Our cloud pressure analysis (see Figure \ref{fig:cloud_pressure}) also reveals a notable shift that accompanies the decrease in cloud activity. A small increase in cloud pressure began during 2019 and continues through 2022, which shows that the clouds during this period of low activity are overall located deeper in the atmosphere than in previous years. While occasionally a few small cloud features appear at mid-latitudes, the south polar region consistently hosted features throughout 2020 - 2022, most notably in the H and K' bands. A notable feature consistent throughout these years is a bright ring of activity near $-65 \degree$ in H band data, which is further explored in \cite{chavez2022}. South Polar Features are also present in F845M data, some of which were identified and tracked in \cite{chavez2022}. This consistent activity highlighted the South Polar region as the primary region of prominent cloud activity during 2020 - 2022. Cloud features near Neptune's south pole lie deeper in the atmosphere than those at mid-latitudes \citep[e.g,][]{gibbard2003,irwin2011,depater2014}. Therefore, the increase in disk-averaged cloud pressure in Figure \ref{fig:cloud_pressure} is most likely caused by the primary region of cloud activity shifting from the mid-latitudes to the south polar region.

\subsection{Connections to Solar Cycle Variations}

As discussed in the Introduction, variations in Neptune's brightness cannot only be caused by seasonal variations. Possible correlations with the solar cycle activity have been proposed and investigated before \citep{lockwoodandthompson1986,hammelandlockwood1997, aplinandharrison2016, roman2022}.  In a recent paper, \cite{roman2022} show that long-term variations in Neptune's mid-infrared intensity, interpreted as being caused by a variation in its stratospheric temperature, may also be related to the Solar cycle. These authors noted the apparent tentative correlation between the stratospheric temperature between 2003 and 2020, the discrete cloud coverage between 1994 and 2011 (as measured by \citet{karkoschka2011clouds}), and the solar Lyman $\alpha$ flux \citep{machol2019lymanalpha}.  Our current analysis of disk-averaged cloud I/F not only shows a consistent trend in cloud coverage, but extends and strengthens the apparent correlation in time.

\begin{figure}[h]
    \centering
    \includegraphics[width=0.95\textwidth]{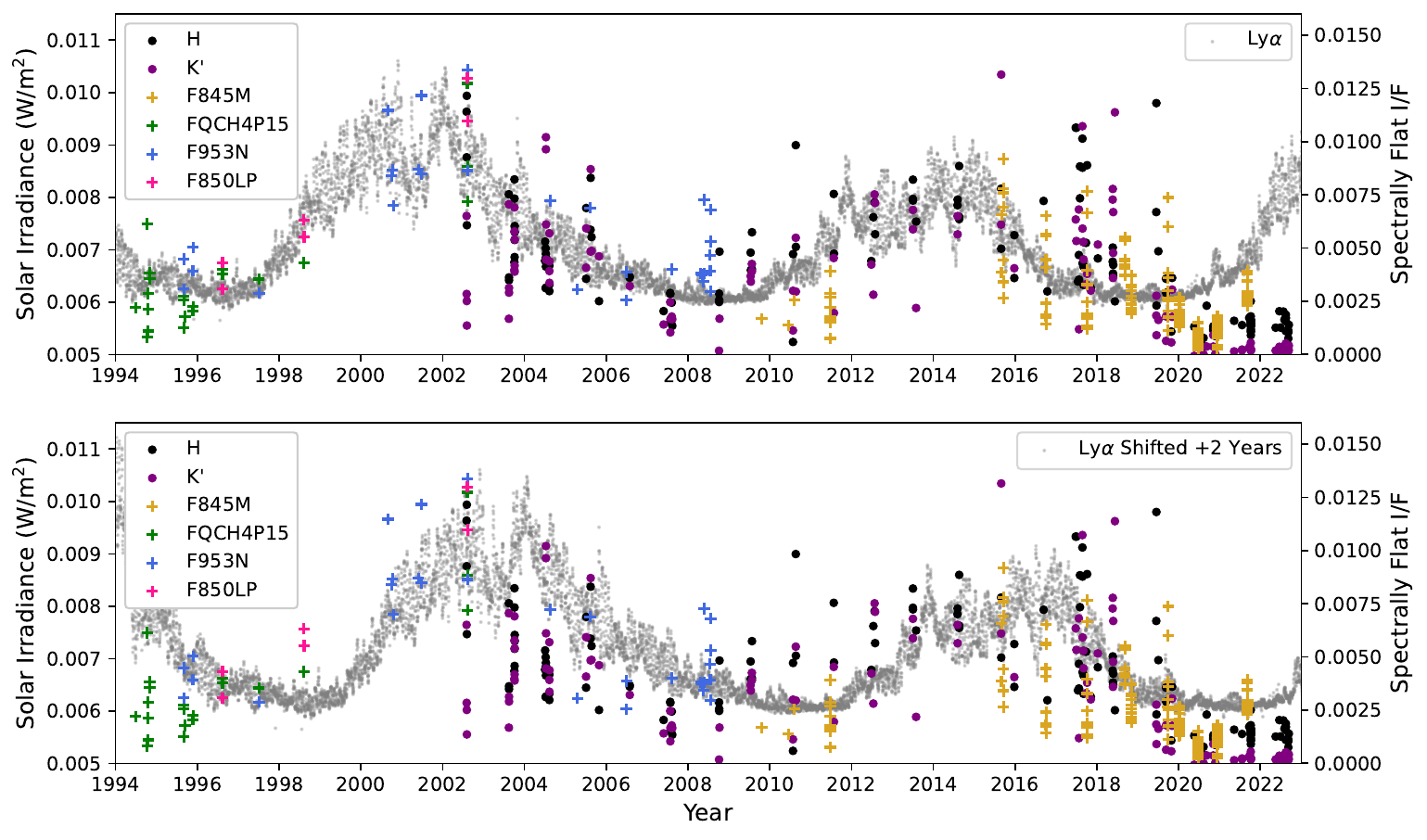}
    \caption{Top: The composite Solar Lyman-alpha irradiance at Earth measured in W/m$^{2}$ from 1994 to 2023 from LISIRD \citep{machol2019lymanalpha}, represented by gray dots. Gaps in measurements were filled in by models derived from solar radio data. Overplotted are the spectrally flat, background-subtracted, disk-averaged brightness results (in units of I/F) shown in Figure \ref{fig:spec_flat_if}. Bottom: The same figure as the top panel, now with the Lyman-alpha emission shifted forward two years.}
    \label{fig:solar_lyman_alpha}
\end{figure}

In Figure \ref{fig:solar_lyman_alpha} we use the Solar Lyman-alpha emission (121.56 nm) measured at Earth as a proxy for the Sun's ultraviolet irradiance and compare it to the cloud variation, following the similar comparison of \cite{roman2022}, but now with additional observations extended over more than two complete solar cycles. The comparison shows similar trends between the solar Lyman-alpha emission and the spectrally flat, disk-averaged cloud I/F (see Figure \ref{fig:spec_flat_if}). Both have a similar periodic pattern characterized by two peaks and three minima. The first peak in the Lyman-alpha emission at $\sim 2002$ was notably stronger than the one around $\sim 2015$. While difficult to determine because of the large scatter in cloud activity around 2017, a similar trend appears to occur in the spectrally flat, disk-averaged cloud I/F, with the first peak appearing slightly stronger than the second. To quantify a possible correlation, we plotted the yearly-averaged Lyman-alpha emission against the spectrally flat I/F in Figure \ref{fig:p_lya_cloud_if}. There is evidence for a linear relationship in this figure, strongly suggesting that solar Lyman-alpha emission is influencing Neptune's cloud activity. We note that the match between the Lyman-alpha and the last minima of the spectrally flat, disk-averaged cloud I/F may be improved with a 2-year shift of the Lyman alpha data forward in time, i.e., a lag time in Neptune's cloud cover by $\sim$2 yr (shown in the bottom panel of Figure \ref{fig:solar_lyman_alpha}). However, the total correlation between solar activity and cloud activity decreases when this is considered (see Figure \ref{fig:p_lya_cloud_if}). The last Lyman-alpha minimum during 2019 is especially notable for how it preceded the drop in cloud activity during late 2019/early 2020.

This correlation indeed suggests that Neptune's cloud cover is affected by the Sun's UV light. A possible explanation for this trend may involve photochemistry. High levels of UV light at wavelengths below $\sim$145 nm will increase the dissociation rate of methane gas, and hence the production of hydrocarbons (C$_x$H$_y$) at high altitudes \citep{moses2018seasonal}. If these hydrocarbons descend in the atmosphere and reach levels where the temperature drops below the condensation temperature of the particular hydrocarbon, they will condense and form hazes and clouds \citep{moses1989neptune,moses1992}. At low levels of UV light, the rate of methane photolysis decreases, and hence such a mechanism may potentially explain the apparent variation in clouds and hazes if the chemical and advection timescales are short enough. 

Interestingly, note that in neither \cite{roman2022} nor our work, maxima in temperature or cloud coverage happened at Neptune's summer solstice, in 2005; both the cloud coverage and temperature were already decreasing, i.e., following the solar activity (UV) cycle. 

In addition to the disk-averaged parameters discussed above, our data show that cloud formation did not diminish in the south polar region. Since the South Polar Features and the bright H-band ring (often not visible in K' band) are located deeper in Neptune's atmosphere, one would indeed expect that solar UV light would have little effect on these features. These, as well as large storms in the atmosphere \citep[such as the one in 2017, discussed in][]{molter2019}, are indicative of convective processes originating deeper in the atmosphere. Such processes will affect the cloud coverage and disk-averaged cloud I/F in unknown ways. It is interesting to note that we sometimes do see clouds at mid-latitudes in 2019-2022, which might be the ``lower-level'' clouds as identified by \cite{depater2014} which presumably rise up from deeper levels. Hence, these clouds should not be affected by solar UV emission. 

The variation in the spatial distribution of clouds in our data and mid-infrared emission in \cite{roman2022} is also intriguing. The south polar region became much brighter in the mid-infrared between 2018 and 2020, in particular in the methane (CH$_4$) and ethane (C$_2$H$_6$) bands \citep{roman2022}. As mentioned by the latter authors, this may indicate either an increased stratospheric temperature and/or an enhanced hydrocarbon abundance. While an increase in methane abundance will heat the atmosphere through absorption of sunlight, an increase in solar UV emission will increase methane photolysis, and hence increase the production of hydrocarbons and hazes. In contrast to the atmospheric heating by methane and hazes, an increase in hydrocarbon emissions would lead to local cooling of the atmosphere, i.e., there is a complicated feedback between photochemical and radiative processes. More detailed modelling of the processing linking clouds, temperatures, and chemistry is needed and will be the focus of future work.  

Despite variations caused by changes in seasons and solar activity, Neptune clearly does also exhibit variability triggered by internal processes. As shown in Figure \ref{fig:p_lya_cloud_if}, the relation between cloud activity and solar activity is not perfectly linear. Intense cloud activity is present during years of decreased solar emissions, therefore internal processes must also play a significant role in modulating cloud activity. The occurrence of dark spots and infrared-bright storms are ``set'' by an internal clock, perhaps much like the quasi-periodic (every 20-30 years) appearance of giant storms in Saturn's atmosphere, which are likely caused by occasional moist convective events, after having been suppressed for decades \citep{liandingersoll2015}. \citet{hammelandlockwood2007} suggested the apparition of large storms in $\sim$1976, 1986 and 1993, with ``transitional'' periods in between. After these events, storms in HST and AO data did not occur until $\sim$2015, and while the large dark spot, NDS2018, was still visible in 2021, no infrared-bright storms were seen after 2019. While the ``transitional'' period characterized by \citet{hammelandlockwood2007} following the bright storms did consist of cloud features filling multiple latitude bands, as indeed occurred from $\sim$2000 to $\sim$2015, the period following the most recent events showed essentially no activity, not even companion clouds to NDS2018 \citep{wong2022}.

\section{Conclusions}

We examined the time variation in the fractional cloud cover and the disk-averaged I/F of Neptune's clouds as derived from near-infrared HST and Keck data between 1994 and 2022. We summarize our findings as follows:

\begin{itemize}
    \item Periodic variation is apparent in Neptune's cloud activity. While it is present in the fractional cloud coverage, it is most notable in the disk-averaged cloud brightness measurements. We documented two cycles of activity with maxima in $\sim 2002$ and 2015, and minima in $\sim 1996$, 2007, and 2020.
    \item A significant transition in Neptune's cloud activity and cloud distribution occurred during late 2019/early 2020 that changed its appearance. This was characterized by the near-disappearance of mid-latitude clouds that were typical throughout previous decades, leaving a primarily blank disk dominated by background hazes. The South Polar region was unaffected by this change and became the primary region of cloud activity, especially in H and K' band. Neptune's new appearance has persisted throughout our most recent data; as of 2022, we have not observed a return of prominent mid-latitudinal cloud activity comparable to earlier years.
    \item The pattern in Neptune's average cloud brightness (i.e., disk-averaged I/F after subtraction of a uniform background atmosphere) shows a correlation with Solar ultraviolet emissions. Our data provide the strongest evidence to date that the discrete cloud coverage appears correlated with the solar cycle, following the findings by \cite{roman2022} and extending the observational record initially reported by \cite{karkoschka2011clouds}.
\end{itemize}

While we documented 2.5 cycles of cloud activity in this paper, more work is necessary to further explore the relation between seasonal changes and solar UV emissions with variations in Neptune's clouds and/or hazes. This relation is, no doubt, complex. For example, an increase in UV sunlight would increase photolysis of methane gas, and hence the production of hydrocarbons and associated hazes. However, an increase in UV sunlight may also lead to a darkening of aerosols and hazes, and hence a decrease in Neptune's overall brightness. Since solar activity is inversely correlated with the intensity of galactic cosmic rays, it may be difficult to disentangle the ion-induced nucleation from aerosol darkening. Stratospheric temperature must affect the haze production rate as well, and this temperature must depend on the season and the stratospheric hydrocarbon production rate (i.e., UV sunlight), while hydrocarbon emissions cool the atmosphere -  i.e., a complex system with feedback mechanisms between photochemical and radiative processes. Finally, processes internal to Neptune must drive infrared-bright "storms", which are likely convective processes perhaps suppressed for decades before making it up to the "surface", while many "storms" appear as companion clouds to Dark Spots, which are vortices located deeper in the atmosphere.
Continued observations of Neptune are also necessary to observe if the new era of diminished cloud activity that began during late 2019/early 2020 will continue in future years.

\section*{Acknowledgments}

We would like to thank the two anonymous referees whose comments greatly improved this manuscript. This work has been supported by the National Science Foundation, NSF Grant AST-1615004 to UC Berkeley. M.Roman was supported by a European Research Council Consolidator Grant (under the European Union's Horizon 2020 research and innovation programme, grant agreement No 723890) at the University of Leicester.

Many of the images were obtained with the W. M. Keck Observatory, which is operated as a scientific partnership among the California Institute of Technology, the University of California and the National Aeronautics and Space Administration. The Observatory was made possible by the generous financial support of the W. M. Keck Foundation.

The authors wish to recognize and acknowledge the very significant cultural role and reverence that the summit of Maunakea has always had within the indigenous Hawaiian community. We are most fortunate to have the opportunity to conduct observations from this mountain.

We further made use of data obtained with the NASA/ESA Hubble Space Telescope, obtained from the data archive at the Space Telescope Science Institute. 

This work used data acquired from the NASA/ESA HST Space Telescope, associated with OPAL program (PI: Simon, GO13937), and archived by the Space Telescope Science Institute, which is operated by the Association of Universities for Research in Astronomy, Inc., under NASA contract NAS 5-26555. All maps are available at \url{http://dx.doi.org/10.17909/T9G593}.

STScI is operated by the Association of Universities for Research in Astronomy, Inc. under NASA contract NAS 5–26555

Research at Lick Observatory is partially supported by a generous gift from Google.

\vspace{5mm}

\software{astropy, matplotlib, nirc2\_reduce, numpy, pandas, scipy}

\appendix

\setcounter{table}{0}
\renewcommand{\thetable}{A\arabic{table}}
\setcounter{figure}{0}
\renewcommand{\thefigure}{A\arabic{figure}}

\begin{figure}[h]
    \centering
    \includegraphics[width=0.95\textwidth]{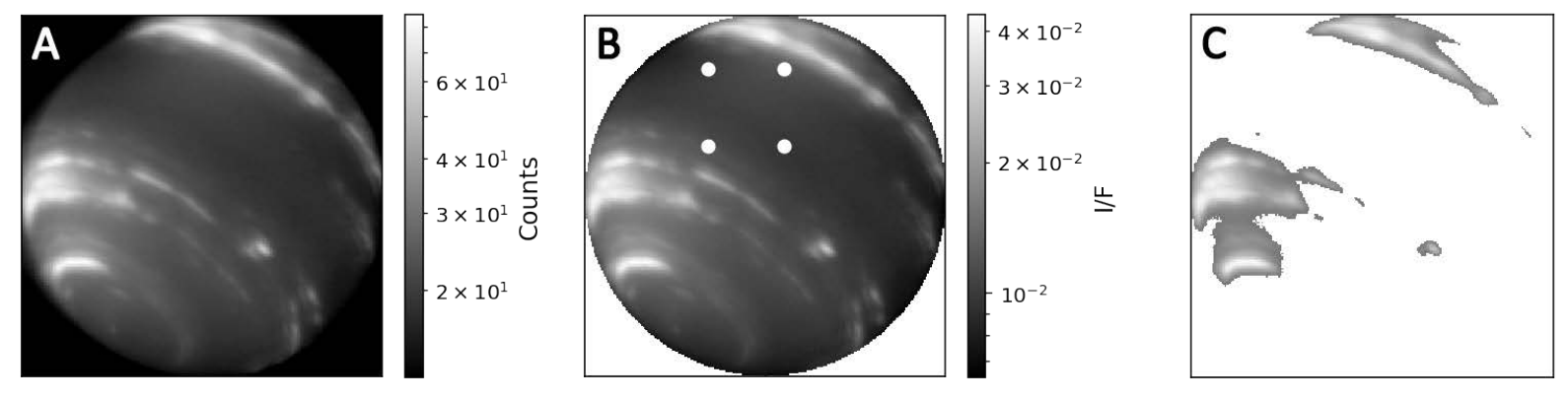}
    \caption{An example of the calibration method described in Section \ref{sec:bootstrap} and of the method used to determine cloud coverage of Neptune's disk. An H band image from September 4, 2017 is shown (displayed logarithmically). Panel A: The uncalibrated image of Neptune. Panel B: The isolated disk after applying our calibration method described in Section \ref{sec:bootstrap})}. The box within the equatorial region used to calculate the background I/F is outlined by four white dots. Panel C: The regions of Neptune's disk that contains clouds are shown (see the method outlined in Section \ref{sec:frac_cloud_cover}).
    \label{fig:bootstrap_box_example}
\end{figure}

Figure \ref{fig:bootstrap_box_example} demonstrates an example of calibrating an image using the method outlined in Section \ref{sec:bootstrap} and shows the cloud regions identified in that image.

\begin{figure}[h]
    \centering
    \includegraphics[width=0.95\textwidth]{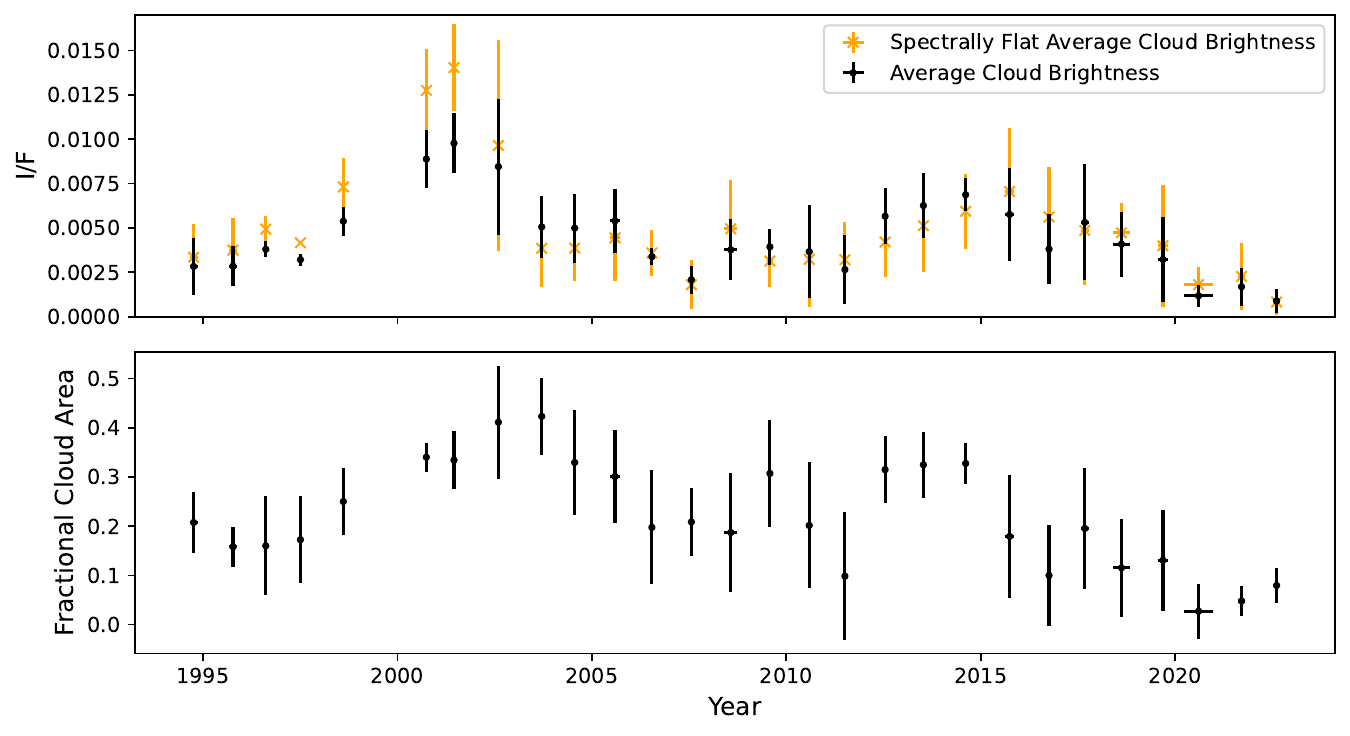}
    \caption{Top: The black points show the spectrally flat, disk-averaged, background-subtracted I/F shown in Figure \ref{fig:spec_flat_if}, now averaged within 1-year bins. The orange points are the disk-averaged, background subtracted I/F from panel D of Figure \ref{fig:if_process}. Bottom: The fractional cloud coverage shown in Figure \ref{fig:cloud_cover} averaged within 1-year bins. The uncertainties for the binned cloud I/F and the binned fractional cloud coverage were determined by taking the standard deviations of the data.\label{fig:binned}}
\end{figure}

\begin{figure}[h]
    \centering
    \includegraphics[width=0.55\textwidth]{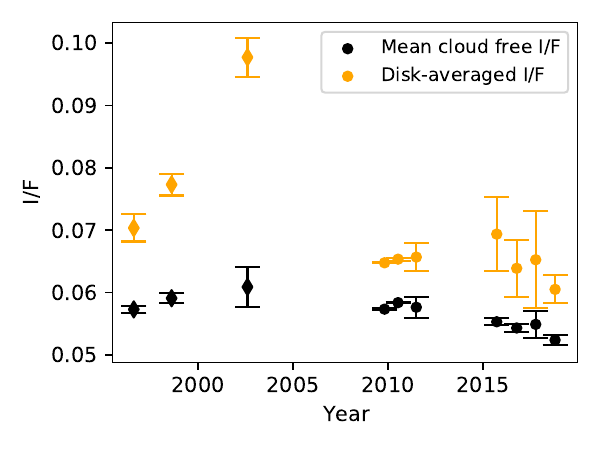}
    \caption{Disk-averaged and mean background I/F for HST F850LP and F845M data. The orange points show the disk-averaged I/F and the black points show the mean background I/F. The symbols distinguish WFPC2 F850LP from 1994 up to 2009 (diamonds) and WFC3 F845M data from 2009 onwards (circles). The error bars show the variance within each year. \citep[Adapted from][]{zorzi2019}.}
    \label{fig:zorzi2019diskvsbackground}
\end{figure}

Here we discuss the sources of uncertainty associated with our determination of the cloud I/F. The three steps were calibrating the non-photometric Keck data, performing background subtraction, and the spectral flattening procedure. 
To characterize the spread in the data, Figure \ref{fig:binned} shows the raw
cloud I/F data from Figure \ref{fig:spec_flat_if}, the spectrally flat cloud I/F data from Figure \ref{fig:spec_flat_if}, and the fractional cloud coverage from Figure \ref{fig:cloud_cover} averaged over 1-year bins. The periodic variation in cloud brightness is present in both the raw and spectrally flat cloud brightness data. This figure also shows how the amount of scatter changes when the average cloud brightness is spectrally flattened. Figure \ref{fig:zorzi2019diskvsbackground} shows the yearly variations in the disk-averaged I/F and the background I/F as measured in F850LP and F845M data.

The uncertainties due to the Keck calibration method described in Section \ref{sec:bootstrap} are shown in Figure \ref{fig:avg_h_kp_background}; these were all equal or less than 0.001 I/F. Uncertainty introduced from background subtraction were determined in Section \ref{sec:bootstrap} by taking the standard deviation of a cloud-free region on Neptune. Typical values can be inferred from the error bars of the gray points on Figure \ref{fig:avg_h_kp_background} for Keck data and from the black points in Figure \ref{fig:zorzi2019diskvsbackground} for HST data, all of which were equal or less than 0.005 I/F.

From Figure \ref{fig:binned} the scatter in the cloud brightness is either unchanged or reduced after the spectral flattening procedure (see orange vs black points). The same periodic pattern in cloud activity can be seen in both cases. This indicates that uncertainty due to the spectral flattening procedure is small in comparison to the other sources. From this, the most significant source of uncertainty is the calibration procedure for non-photometric Keck data done in Section \ref{sec:bootstrap}, with background subtraction being the next largest contributor. We expect up to 20\% uncertainty to the cloud I/F results that were determined from non-photometric Keck data, and up to 10\% uncertainty in the results determined from all other data.  These error bars remain smaller than the 90$\%$ amplitude of the periodic trend we observe. We conclude that the periodic variation seen over the 28-years of data is caused primarily by cloud activity, and not error introduced by our methodology.

\begin{figure}[h]
    \centering
    \includegraphics[width=\textwidth]{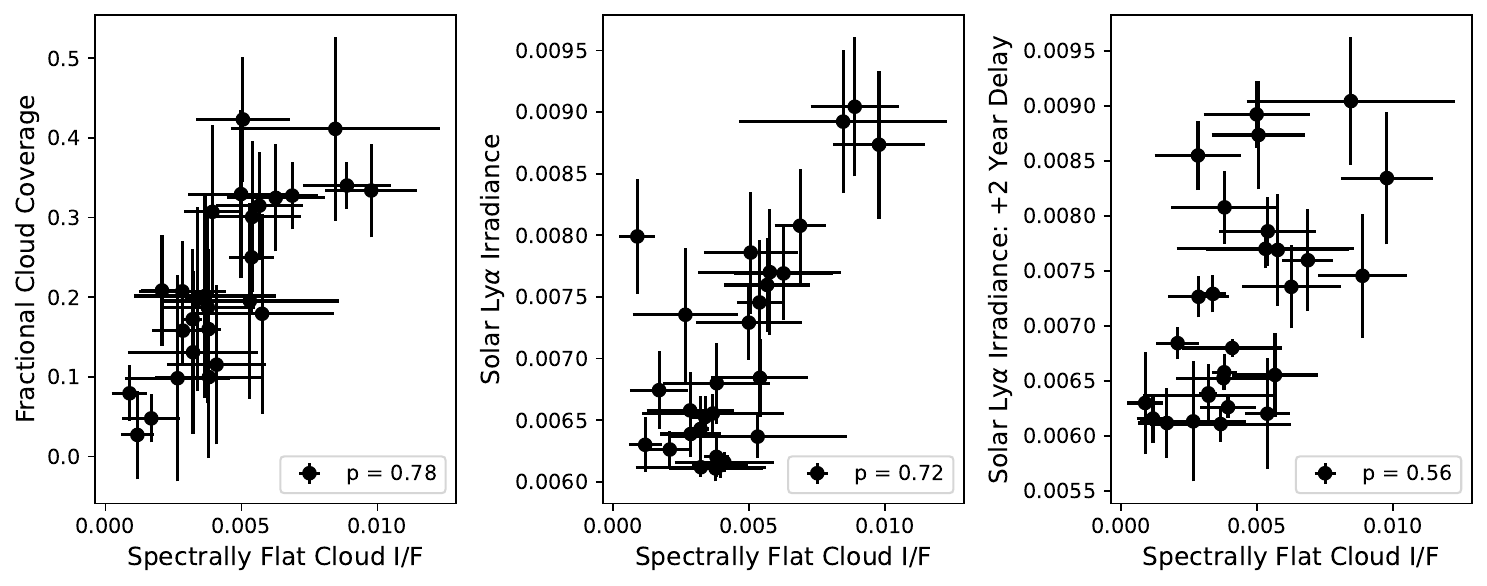}
    \caption{Left: The binned spectrally flat cloud I/F plotted against the binned fractional cloud cover as measured in HST and Keck data (see black points in Figure \ref{fig:binned}). The Pearson correlation coefficient is 0.78. Center: The yearly-averaged Lyman-Alpha solar irradiance plotted against the yearly-averaged spectrally flat cloud I/F. The Pearson correlation is 0.72. Right: The yearly-averaged Lyman-Alpha solar irradiance with a $+2$ year time delay plotted against the yearly-averaged spectrally flat cloud I/F. The Pearson correlation coefficient is 0.56, showing a weaker correlation with cloud activity when compared with the original solar Ly$\alpha$ emisison.}
    \label{fig:p_lya_cloud_if}
\end{figure}

Figure \ref{fig:p_lya_cloud_if} shows the correlations between the fractional cloud coverage and the Lyman-alpha emission with the spectrally flat cloud I/F. All quantities shown are averaged over yearly bins, and the uncertainties are determined by the standard deviations of the data in the bins. The Pearson correlation coefficient is used to characterize the correlation; numbers closer to $+1$ indicate a stronger positive correlation. The left panel of Figure \ref{fig:p_lya_cloud_if} shows a strong correlation (p=0.78) between the cloud cover and the average cloud brightness, confirming that they can both be used as measurements of cloud activity. In addition, a positive correlation between the average cloud brightness and the solar Ly$\alpha$ emission is present (p=0.72). When a 2 year delay in the solar Ly$\alpha$ emission is introduced (right panel of Figure \ref{fig:p_lya_cloud_if}), the correlation between solar activity and cloud activity becomes weaker, but it is still positively correlated (p=0.56). This suggests that there may not be an overall delay in cloud activity in comparison to solar activity. In both cases, we show that a positive correlation exists between cloud activity and solar emission.

Figures \ref{fig:hst_pages_of_images}, \ref{fig:keck_pages_of_images}, and \ref{fig:lick_pages_of_images} show the images used in this paper from the Keck Observatory, Hubble Space Telescope, and Lick Observatory, respectively. Neptune's north pole is pointing roughly upwards in all images.

\begin{figure}
    \centering
    \includegraphics[width=0.99\textwidth]{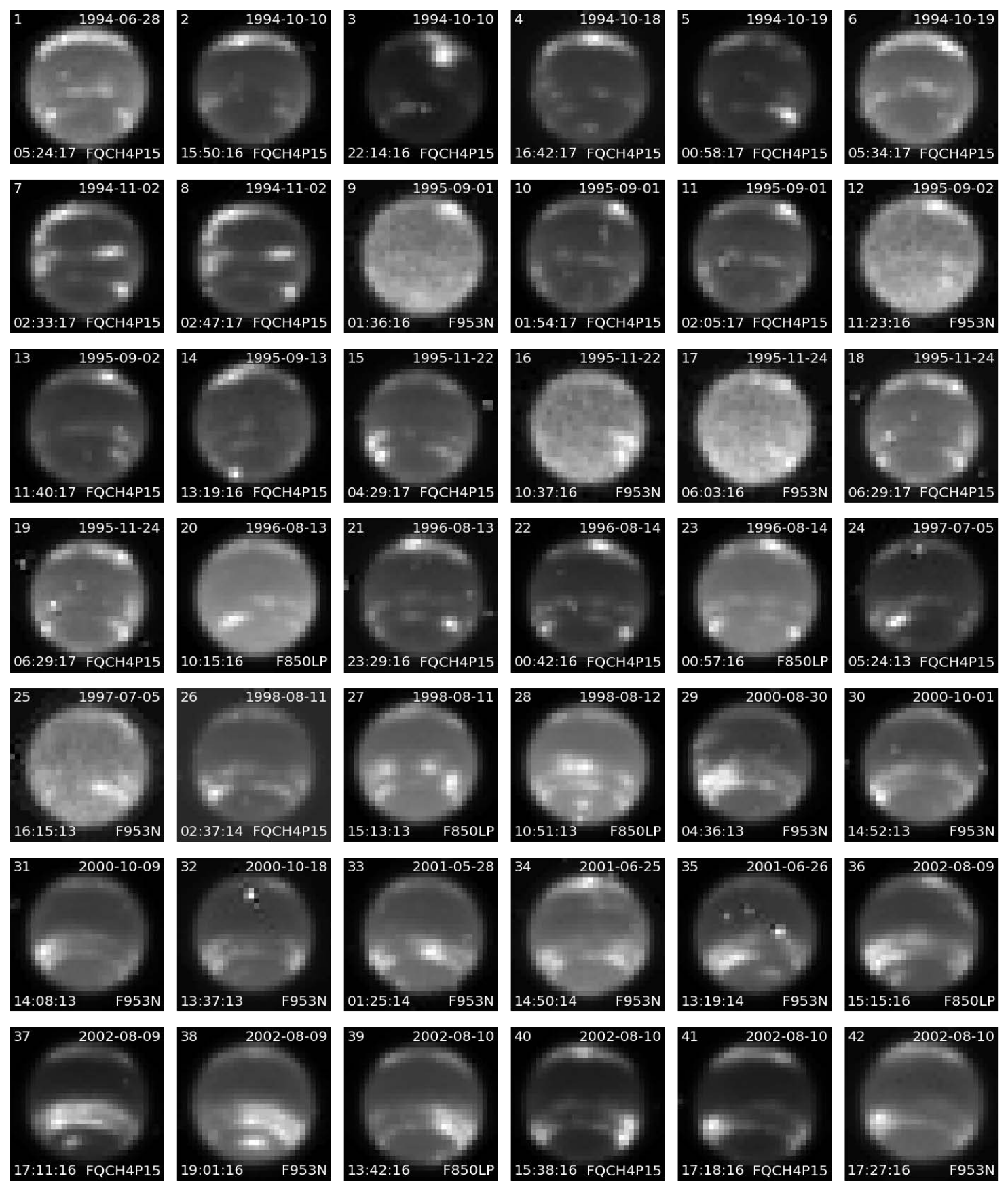}
\end{figure}

\begin{figure}
    \centering
    \includegraphics[width=0.99\textwidth]{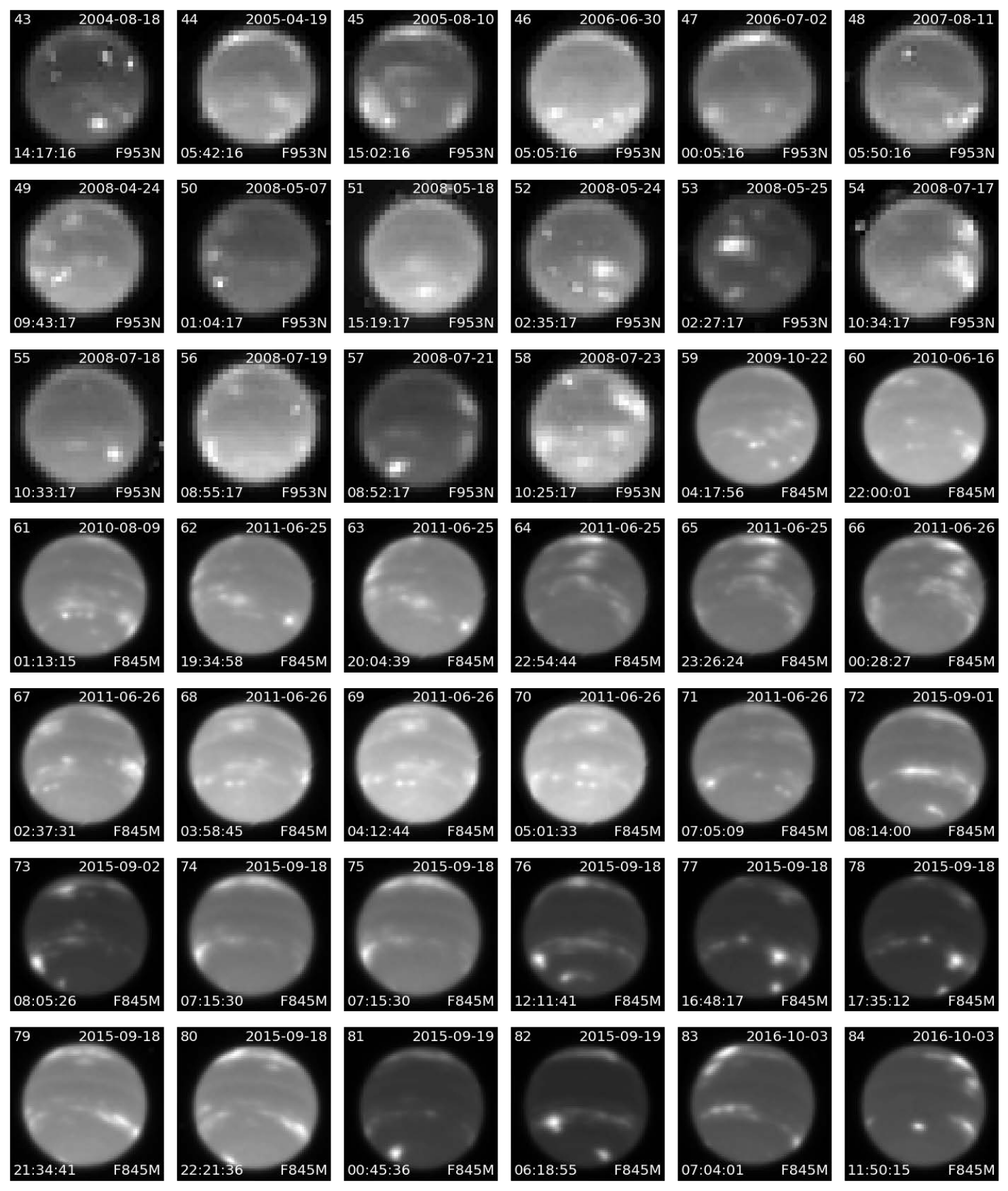}
\end{figure}

\begin{figure}
    \centering
    \includegraphics[width=0.99\textwidth]{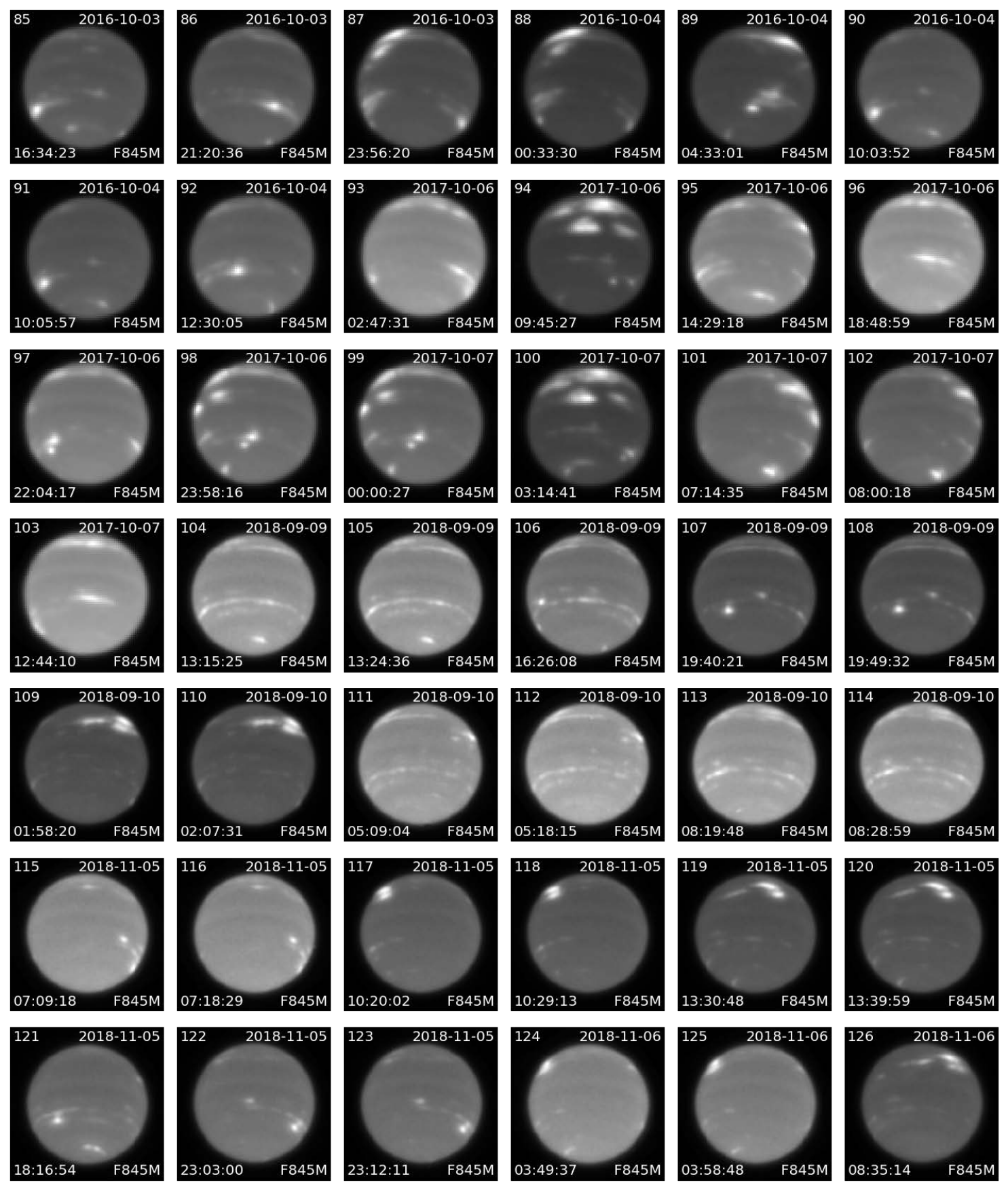}
\end{figure}
    
\begin{figure}
    \centering
    \includegraphics[width=0.99\textwidth]{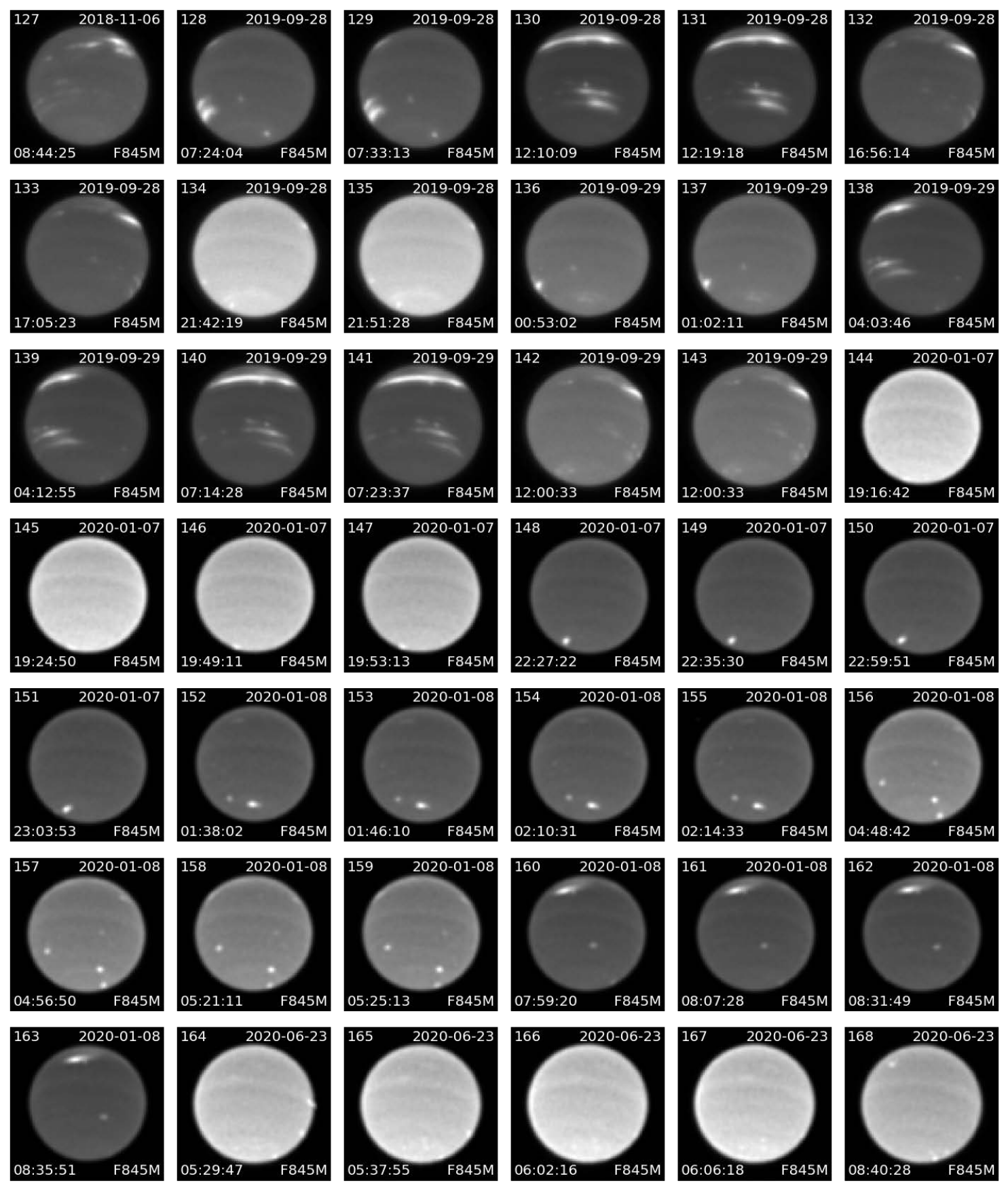}
\end{figure}

\begin{figure}
    \centering
    \includegraphics[width=0.99\textwidth]{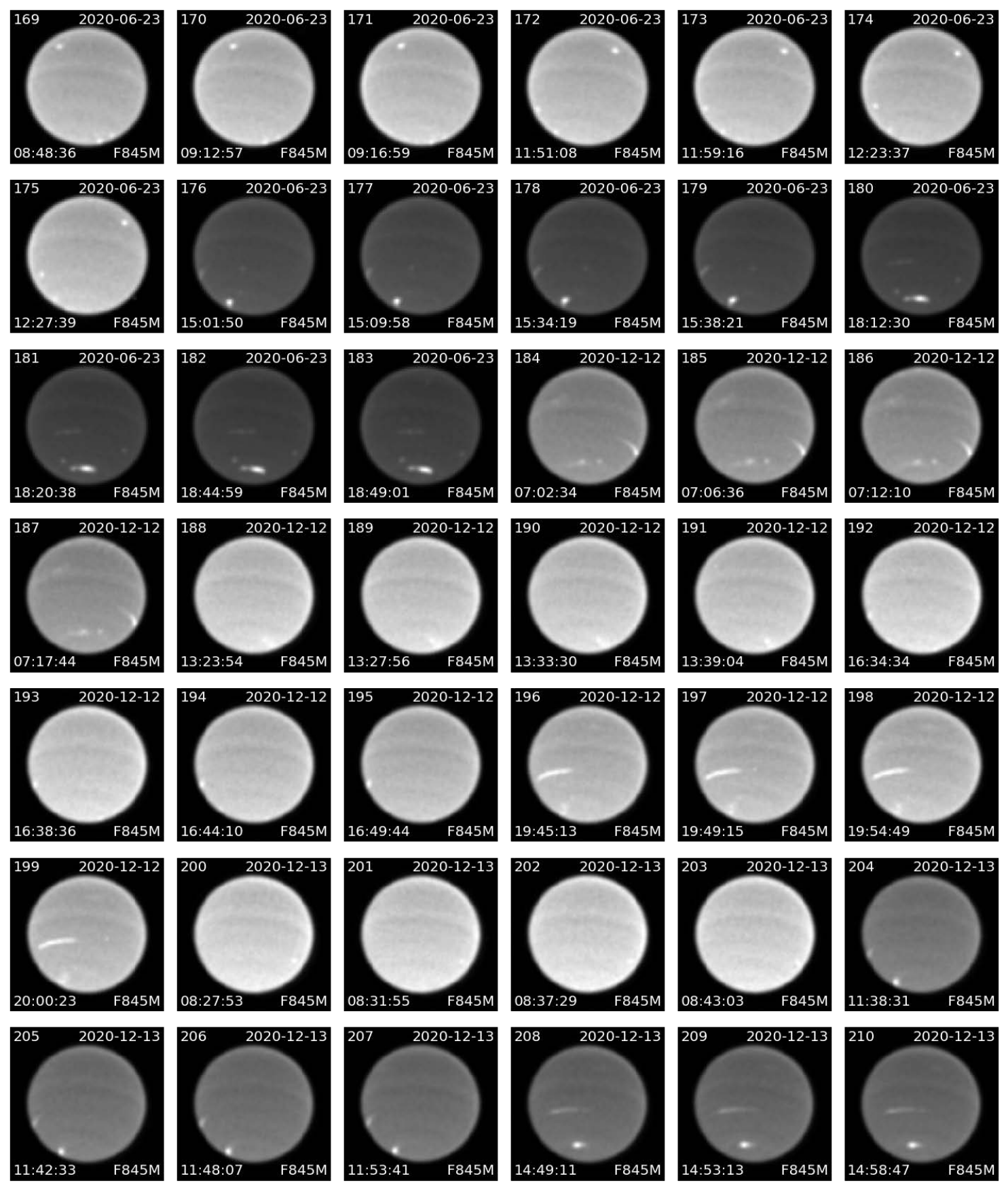}
\end{figure}

\begin{figure}
    \centering
    \includegraphics[width=0.99\textwidth]{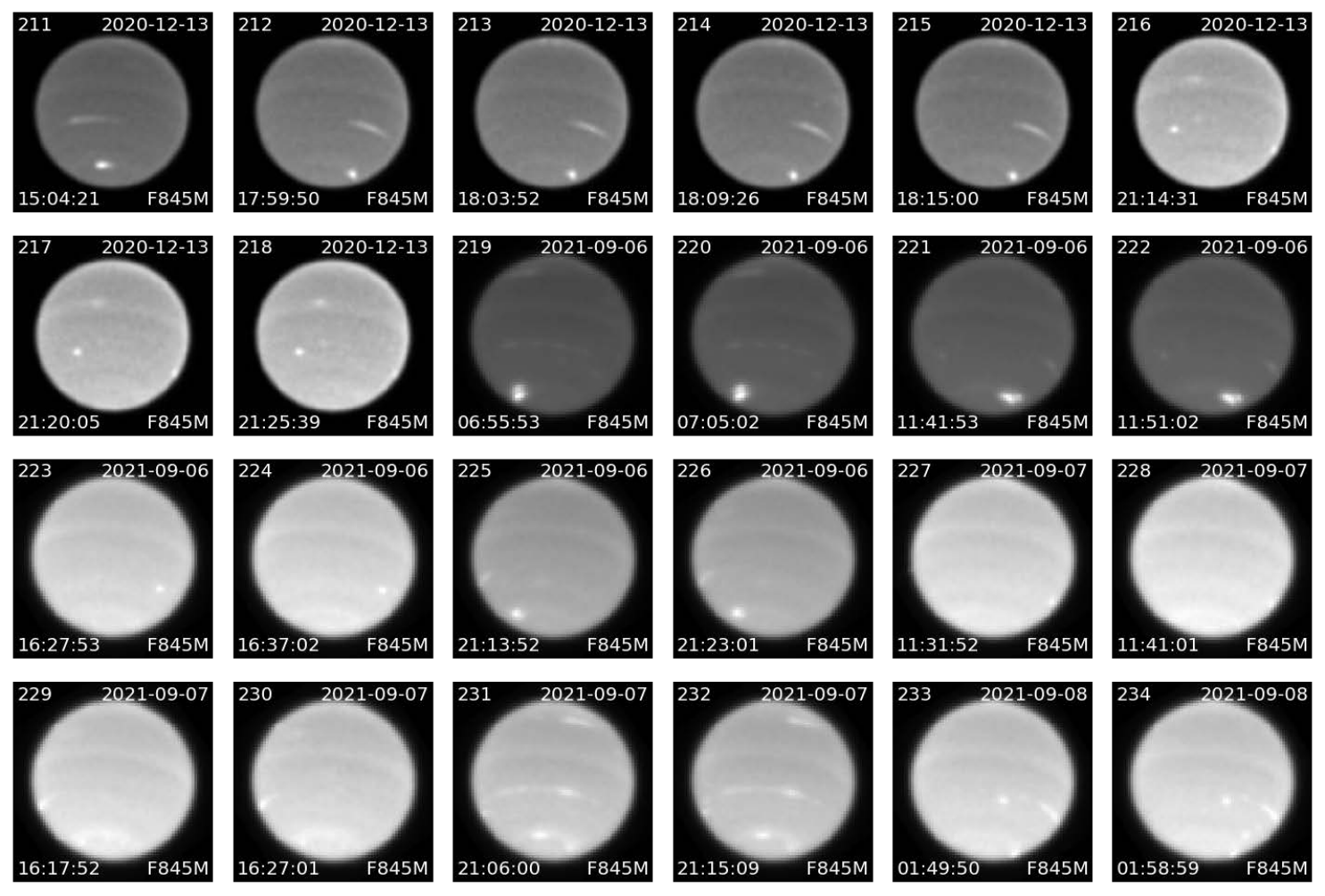}
    \caption{HST data from the WFPC2 (FQCH4P15, F953N, and F850LP) between 1994 and 2009 and from WFC3 (F845M) between 2009 and 2021. Images are displayed to best display the contrast between cloud features and background hazes.}
    \label{fig:hst_pages_of_images}
\end{figure}

\clearpage

\begin{figure}
    \centering
    \includegraphics[width=0.99\textwidth]{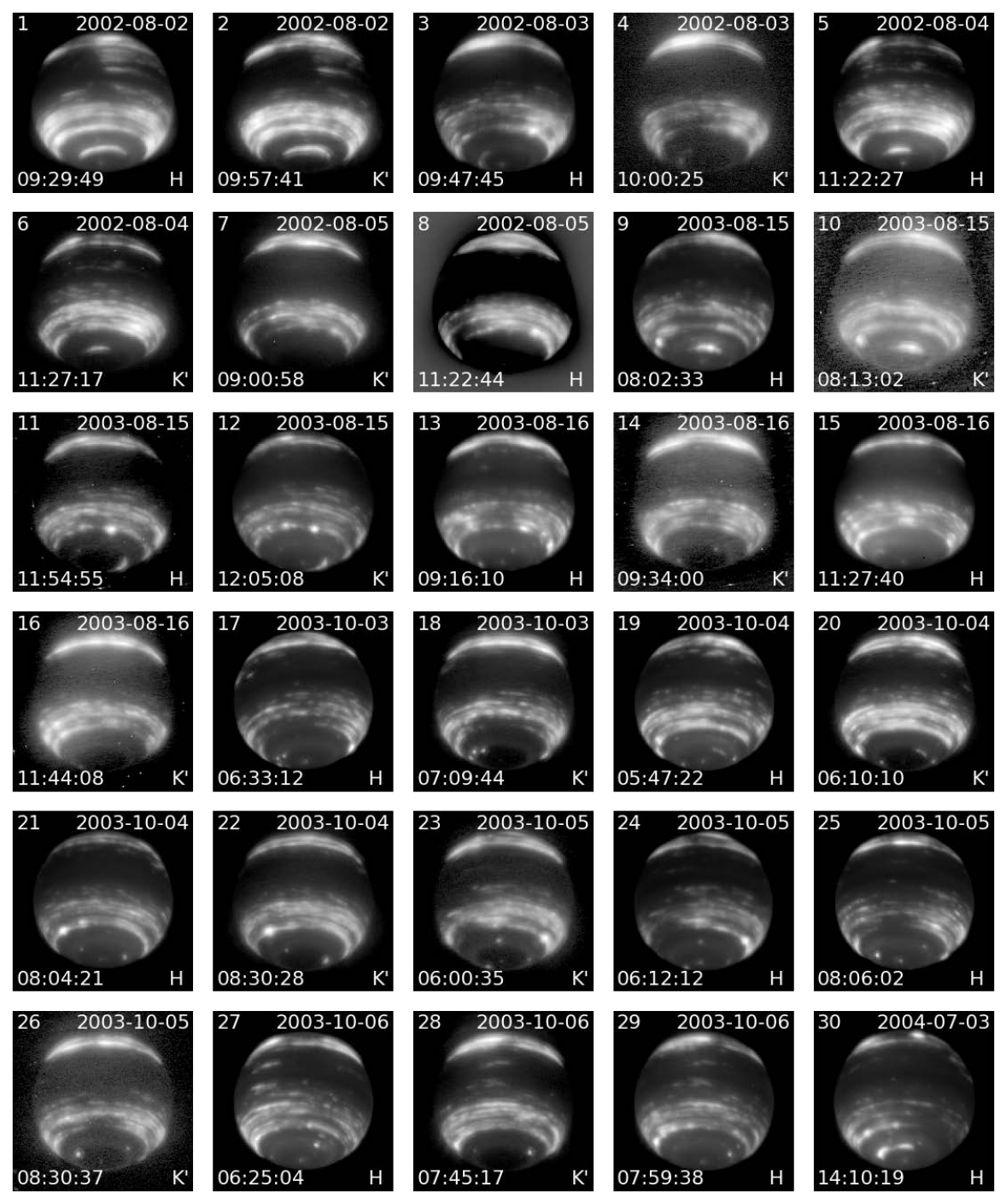}
\end{figure}

\begin{figure}
    \centering
    \includegraphics[width=0.99\textwidth]{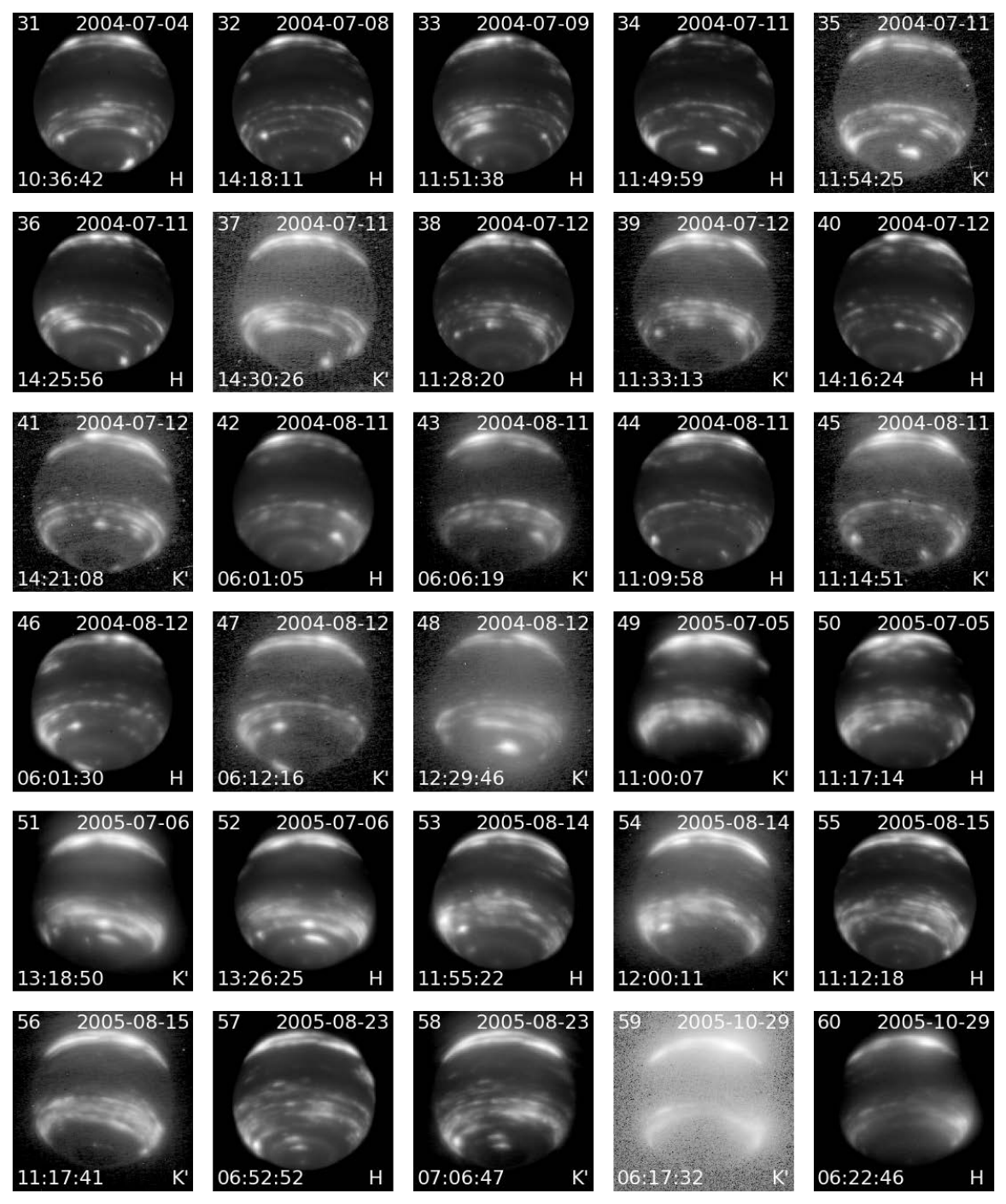}
\end{figure}

\begin{figure}
    \centering
    \includegraphics[width=0.99\textwidth]{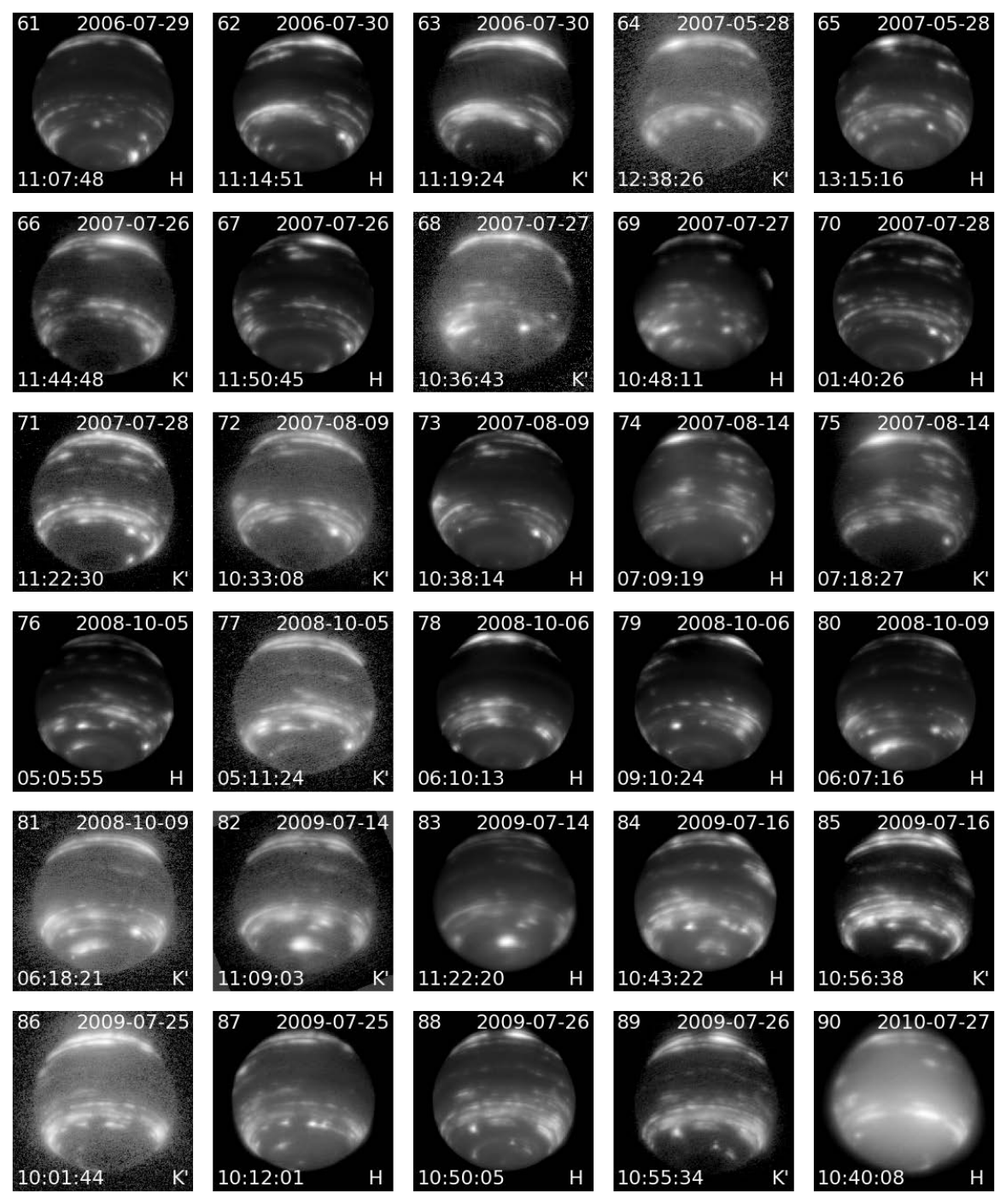}
\end{figure}

\begin{figure}
    \centering
    \includegraphics[width=0.99\textwidth]{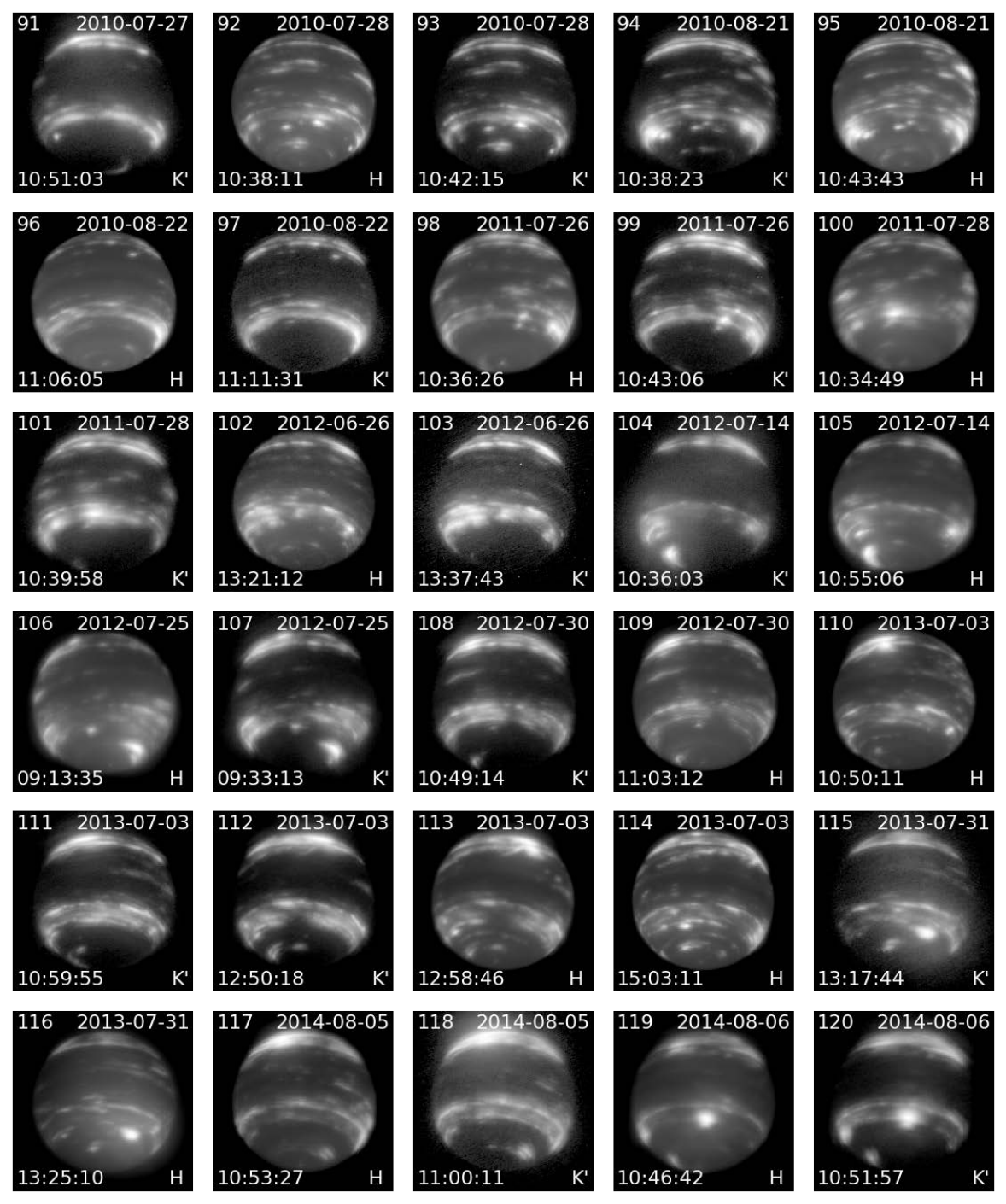}
\end{figure}

\begin{figure}
    \centering
    \includegraphics[width=0.99\textwidth]{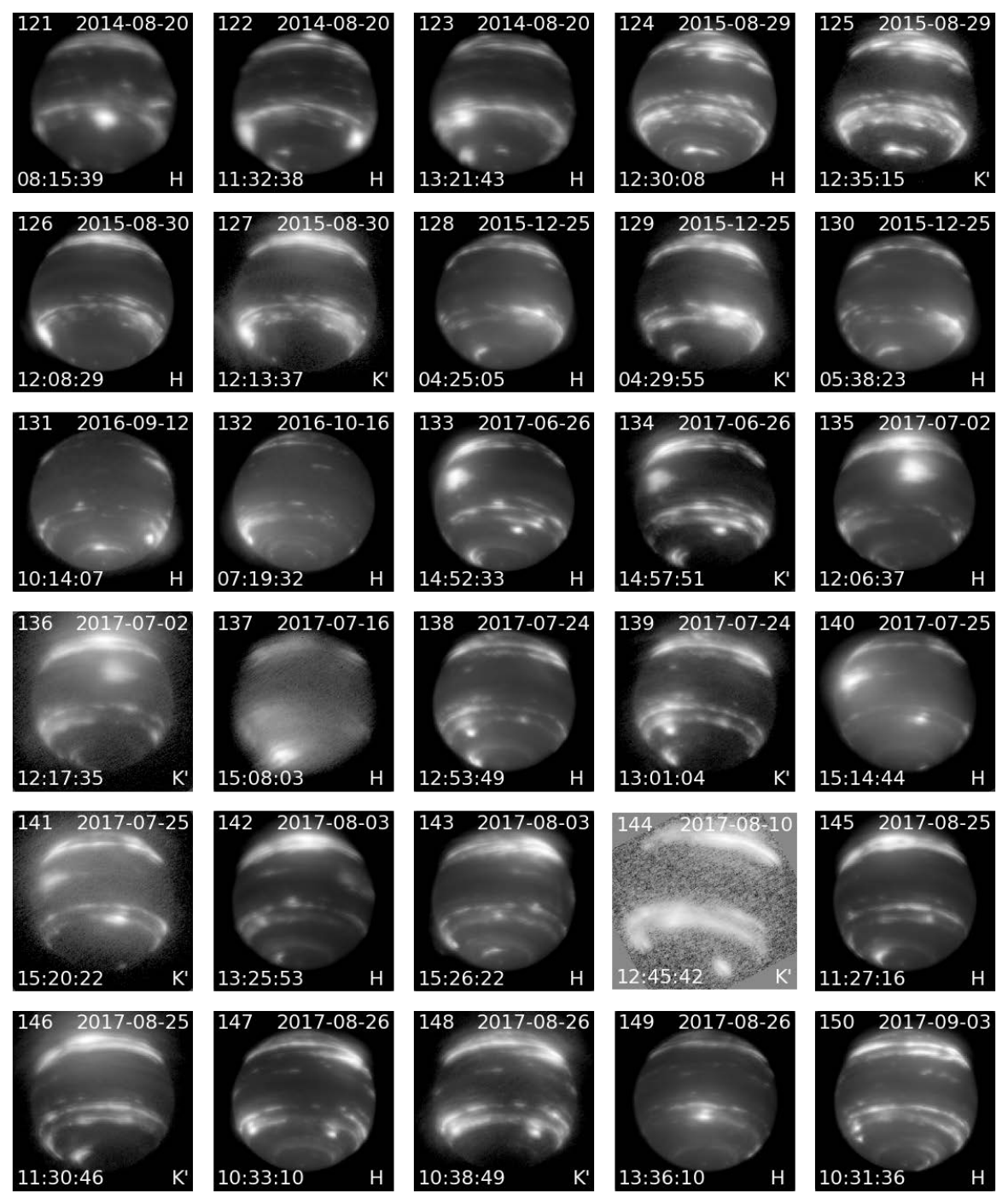}
\end{figure}

\begin{figure}
    \centering
    \includegraphics[width=0.99\textwidth]{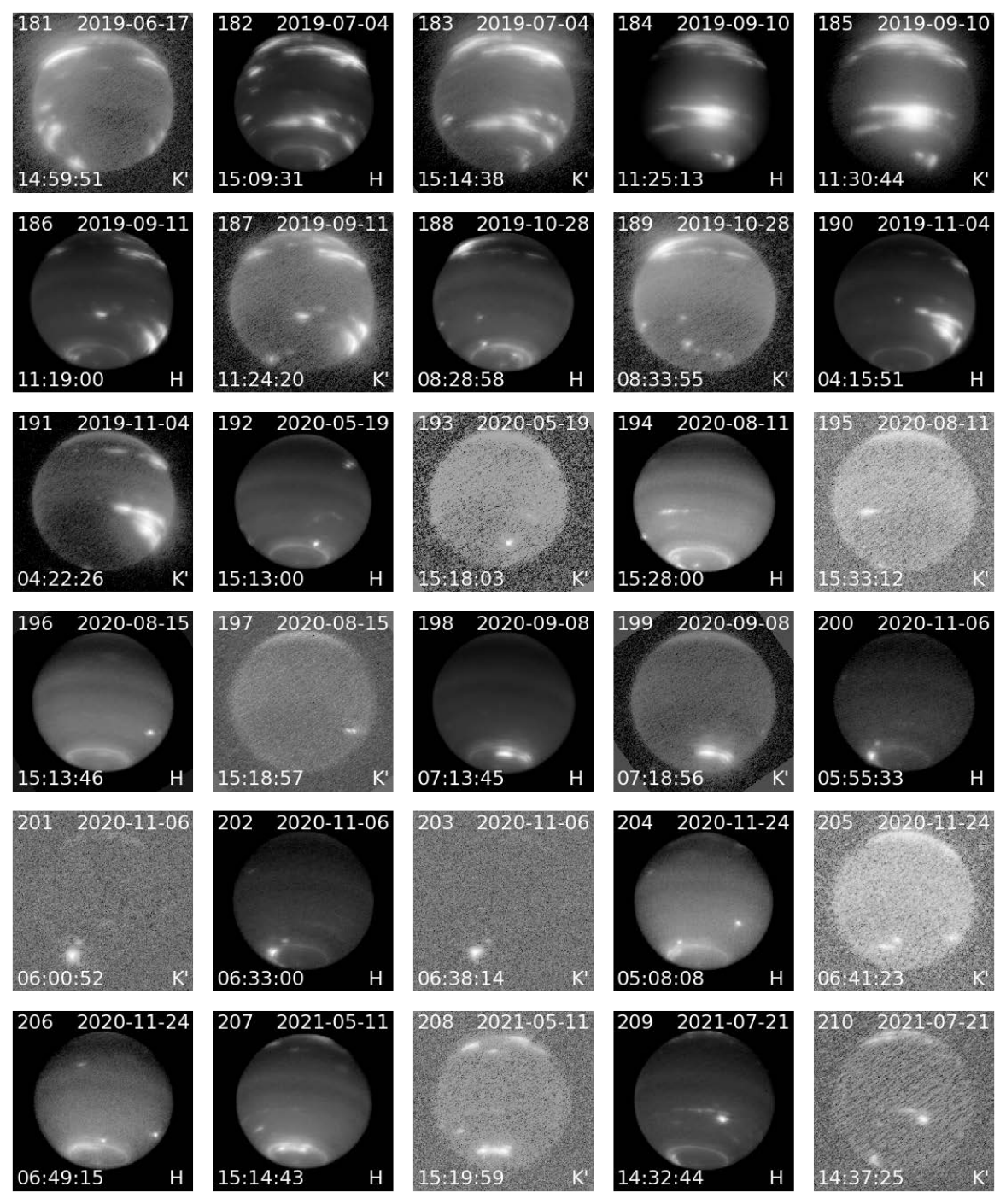}
\end{figure}

\begin{figure}
    \centering
    \includegraphics[width=0.99\textwidth]{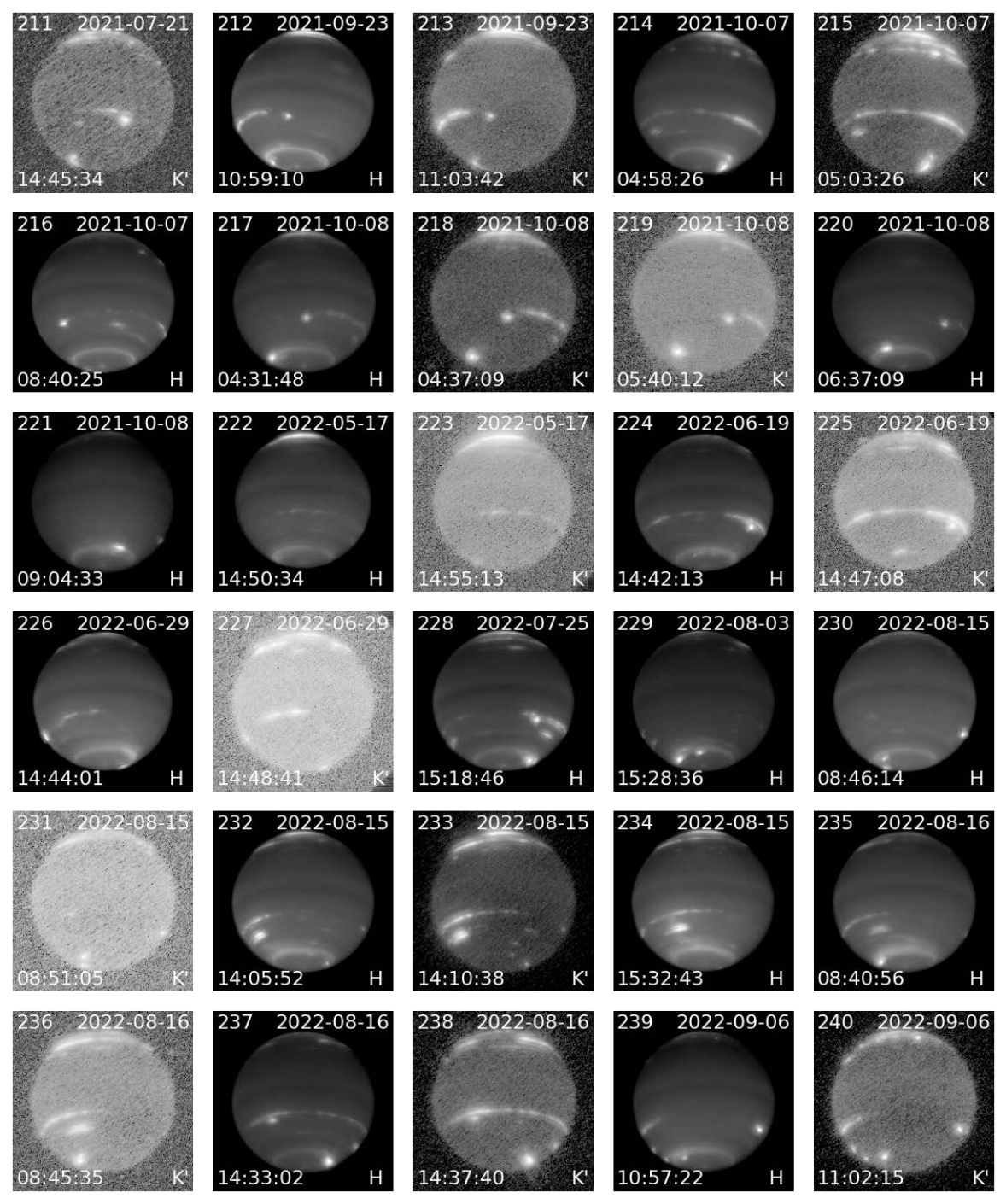}
\end{figure}

\begin{figure}
    \centering
    \includegraphics[width=0.99\textwidth]{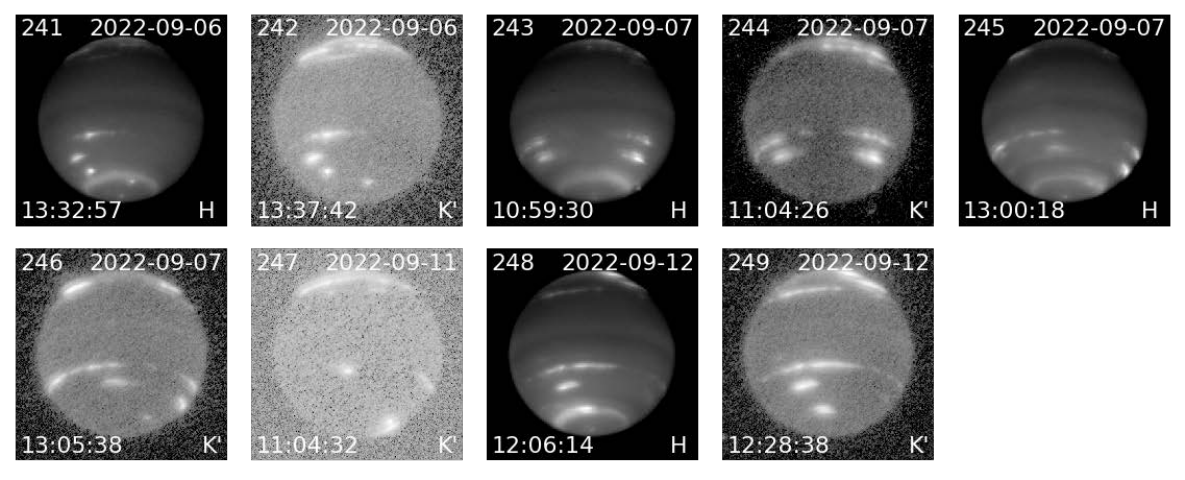}
    \caption{Select Keck images (H and K' band) between 2002 and 2021 used in this paper shown on a logarithmic scale. Images within the same filter taken minutes apart have been excluded. Images are displayed to best display the contrast between cloud features and background hazes. The image stretch for 2020 K' band images was intentionally narrowed to better display Neptune's faint disk.}
    \label{fig:keck_pages_of_images}
\end{figure}

\begin{figure}
    \centering
    \includegraphics[width=0.99\textwidth]{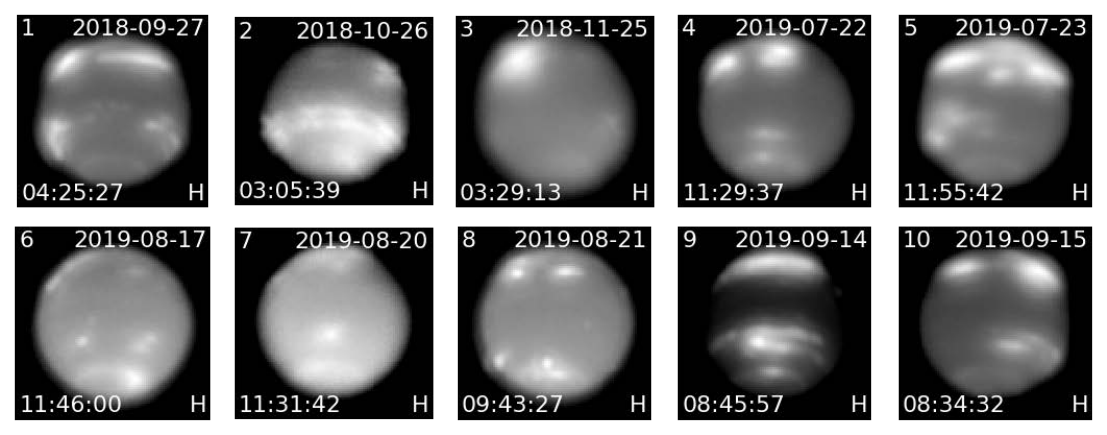}
    \caption{Lick H band images from 2018 and 2019 used in this paper. Images are displayed logarithmically to best display the contrast between cloud features and background hazes.}
    \label{fig:lick_pages_of_images}
\end{figure}

\startlongtable
\begin{deluxetable}{ccccc}
\tablecaption{HST Data Used in this Paper\label{tab:hst_data_table}}
\tablewidth{0pt}
\tablehead{
\colhead{Date} & \colhead{Time} & \colhead{Filter} & \colhead{Program ID} & \colhead{PI}
}
\startdata
1994-06-28 & 05:24:17 & FQCH4P15 & 5221 & Trauger \\
1994-10-10 & 15:50:16 & FQCH4P15 & 5329 & Hammel \\
1994-10-10 & 22:14:16 & FQCH4P15 & 5329 & Hammel \\
1994-10-18 & 16:42:17 & FQCH4P15 & 5329 & Hammel \\
1994-10-19 & 00:58:17 & FQCH4P15 & 5329 & Hammel \\
1994-10-19 & 05:34:17 & FQCH4P15 & 5329 & Hammel \\
1994-11-02 & 02:33:17 & FQCH4P15 & 5329 & Hammel \\
1994-11-02 & 02:47:17 & FQCH4P15 & 5329 & Hammel \\
1995-09-01 & 01:36:16 & F953N & 5831 & Hammel \\
1995-09-01 & 01:54:17 & FQCH4P15 & 5831 & Hammel \\
1995-09-01 & 02:05:17 & FQCH4P15 & 5831 & Hammel \\
1995-09-02 & 11:23:16 & F953N & 5831 & Hammel \\
1995-09-02 & 11:40:17 & FQCH4P15 & 5831 & Hammel \\
1995-09-13 & 13:19:16 & FQCH4P15 & 6219 & Trauger \\
1995-11-22 & 04:29:17 & FQCH4P15 & 5831 & Hammel \\
1995-11-22 & 10:37:16 & F953N & 5831 & Hammel \\
1995-11-24 & 06:03:16 & F953N & 5831 & Hammel \\
1995-11-24 & 06:29:17 & FQCH4P15 & 5831 & Hammel \\
1995-11-24 & 06:29:17 & FQCH4P15 & 5831 & Hammel \\
1996-08-13 & 10:15:16 & F850LP & 6650 & Sromovsky \\
1996-08-13 & 23:29:16 & FQCH4P15 & 6650 & Sromovsky \\
1996-08-14 & 00:42:16 & FQCH4P15 & 6650 & Sromovsky \\
1996-08-14 & 00:57:16 & F850LP & 6650 & Sromovsky \\
1997-07-05 & 05:24:13 & FQCH4P15 & 5831 & Hammel \\
1997-07-05 & 16:15:13 & F953N & 5831 & Hammel \\
1998-08-11 & 02:37:14 & FQCH4P15 & 7324 & Sromovsky \\
1998-08-11 & 15:13:13 & F850LP & 7324 & Sromovsky \\
1998-08-12 & 10:51:13 & F850LP & 7324 & Sromovsky \\
2000-08-30 & 04:36:13 & F953N & 8634 & Rages \\
2000-10-01 & 14:52:13 & F953N & 8634 & Rages \\
2000-10-09 & 14:08:13 & F953N & 8634 & Rages \\
2000-10-18 & 13:37:13 & F953N & 8634 & Rages \\
2001-05-28 & 01:25:14 & F953N & 8634 & Rages \\
2001-06-25 & 14:50:14 & F953N & 8634 & Rages \\
2001-06-26 & 13:19:14 & F953N & 8634 & Rages \\
2002-08-09 & 15:15:16 & F850LP & 9393 & Sromovsky \\
2002-08-09 & 17:11:16 & FQCH4P15 & 9393 & Sromovsky \\
2002-08-09 & 19:01:16 & F953N & 9393 & Sromovsky \\
2002-08-10 & 13:42:16 & F850LP & 9393 & Sromovsky \\
2002-08-10 & 15:38:16 & FQCH4P15 & 9393 & Sromovsky \\
2002-08-10 & 17:18:16 & FQCH4P15 & 9393 & Sromovsky \\
2002-08-10 & 17:27:16 & F953N & 9393 & Sromovsky \\
2004-08-18 & 14:17:16 & F953N & 10170 & Rages \\
2005-04-19 & 05:42:16 & F953N & 10170 & Rages \\
2005-08-10 & 15:02:16 & F953N & 10534 & Rages \\
2006-06-30 & 05:05:16 & F953N & 10534 & Rages \\
2006-07-02 & 00:05:16 & F953N & 10534 & Rages \\
2007-08-11 & 05:50:16 & F953N & 11156 & Rages \\
2008-04-24 & 09:43:17 & F953N & 11156 & Rages \\
2008-05-07 & 01:04:17 & F953N & 11156 & Rages \\
2008-05-18 & 15:19:17 & F953N & 11156 & Rages \\
2008-05-24 & 02:35:17 & F953N & 11156 & Rages \\
2008-05-25 & 02:27:17 & F953N & 11156 & Rages \\
2008-07-17 & 10:34:17 & F953N & 11156 & Rages \\
2008-07-18 & 10:33:17 & F953N & 11156 & Rages \\
2008-07-19 & 08:55:17 & F953N & 11156 & Rages \\
2008-07-21 & 08:52:17 & F953N & 11156 & Rages \\
2008-07-23 & 10:25:17 & F953N & 11156 & Rages \\
2009-10-22 & 04:17:56 & F845M & 11630 & Rages \\
2010-06-16 & 22:00:01 & F845M & 11630 & Rages \\
2010-08-09 & 01:13:15 & F845M & 11630 & Rages \\
2011-06-25 & 19:34:58 & F845M & 12675 & Noll \\
2011-06-25 & 20:04:39 & F845M & 12675 & Noll \\
2011-06-25 & 22:54:44 & F845M & 12675 & Noll \\
2011-06-25 & 23:26:24 & F845M & 12675 & Noll \\
2011-06-26 & 00:28:27 & F845M & 12675 & Noll \\
2011-06-26 & 02:37:31 & F845M & 12675 & Noll \\
2011-06-26 & 03:58:45 & F845M & 12675 & Noll \\
2011-06-26 & 04:12:44 & F845M & 12675 & Noll \\
2011-06-26 & 05:01:33 & F845M & 12675 & Noll \\
2011-06-26 & 07:05:09 & F845M & 12675 & Noll \\
2015-09-01 & 08:14:00 & F845M & 14044 & de Pater \\
2015-09-02 & 08:05:26 & F845M & 14044 & de Pater \\
2015-09-18 & 07:15:30 & F845M & 13937 & Simon \\
2015-09-18 & 07:15:30 & F845M & 13937 & Simon \\
2015-09-18 & 12:11:41 & F845M & 13937 & Simon \\
2015-09-18 & 16:48:17 & F845M & 13937 & Simon \\
2015-09-18 & 17:35:12 & F845M & 13937 & Simon \\
2015-09-18 & 21:34:41 & F845M & 13937 & Simon \\
2015-09-18 & 22:21:36 & F845M & 13937 & Simon \\
2015-09-19 & 00:45:36 & F845M & 13937 & Simon \\
2015-09-19 & 06:18:55 & F845M & 13937 & Simon \\
2016-10-03 & 07:04:01 & F845M & 14334 & Simon \\
2016-10-03 & 11:50:15 & F845M & 14334 & Simon \\
2016-10-03 & 16:34:23 & F845M & 14334 & Simon \\
2016-10-03 & 21:20:36 & F845M & 14334 & Simon \\
2016-10-03 & 23:56:20 & F845M & 14334 & Simon \\
2016-10-04 & 00:33:30 & F845M & 14334 & Simon \\
2016-10-04 & 04:33:01 & F845M & 14334 & Simon \\
2016-10-04 & 10:03:52 & F845M & 14334 & Simon \\
2016-10-04 & 10:05:57 & F845M & 14334 & Simon \\
2016-10-04 & 12:30:05 & F845M & 14334 & Simon \\
2017-10-06 & 02:47:31 & F845M & 14756 & Simon \\
2017-10-06 & 09:45:27 & F845M & 14756 & Simon \\
2017-10-06 & 14:29:18 & F845M & 14756 & Simon \\
2017-10-06 & 18:48:59 & F845M & 14756 & Simon \\
2017-10-06 & 22:04:17 & F845M & 14756 & Simon \\
2017-10-06 & 23:58:16 & F845M & 14756 & Simon \\
2017-10-07 & 00:00:27 & F845M & 14756 & Simon \\
2017-10-07 & 03:14:41 & F845M & 14756 & Simon \\
2017-10-07 & 07:14:35 & F845M & 14756 & Simon \\
2017-10-07 & 08:00:18 & F845M & 14756 & Simon \\
2017-10-07 & 12:44:10 & F845M & 14756 & Simon \\
2018-09-09 & 13:15:25 & F845M & 15262 & Simon \\
2018-09-09 & 13:24:36 & F845M & 15262 & Simon \\
2018-09-09 & 16:26:08 & F845M & 15262 & Simon \\
2018-09-09 & 19:40:21 & F845M & 15262 & Simon \\
2018-09-09 & 19:49:32 & F845M & 15262 & Simon \\
2018-09-10 & 01:58:20 & F845M & 15262 & Simon \\
2018-09-10 & 02:07:31 & F845M & 15262 & Simon \\
2018-09-10 & 05:09:04 & F845M & 15262 & Simon \\
2018-09-10 & 05:18:15 & F845M & 15262 & Simon \\
2018-09-10 & 08:19:48 & F845M & 15262 & Simon \\
2018-09-10 & 08:28:59 & F845M & 15262 & Simon \\
2018-11-05 & 07:09:18 & F845M & 15262 & Simon \\
2018-11-05 & 07:18:29 & F845M & 15262 & Simon \\
2018-11-05 & 10:20:02 & F845M & 15262 & Simon \\
2018-11-05 & 10:29:13 & F845M & 15262 & Simon \\
2018-11-05 & 13:30:48 & F845M & 15262 & Simon \\
2018-11-05 & 13:39:59 & F845M & 15262 & Simon \\
2018-11-05 & 18:16:54 & F845M & 15262 & Simon \\
2018-11-05 & 23:03:00 & F845M & 15262 & Simon \\
2018-11-05 & 23:12:11 & F845M & 15262 & Simon \\
2018-11-06 & 03:49:37 & F845M & 15262 & Simon \\
2018-11-06 & 03:58:48 & F845M & 15262 & Simon \\
2018-11-06 & 08:35:14 & F845M & 15262 & Simon \\
2018-11-06 & 08:44:25 & F845M & 15262 & Simon \\
2019-09-28 & 07:24:04 & F845M & 15502 & Simon \\
2019-09-28 & 07:33:13 & F845M & 15502 & Simon \\
2019-09-28 & 12:10:09 & F845M & 15502 & Simon \\
2019-09-28 & 12:19:18 & F845M & 15502 & Simon \\
2019-09-28 & 16:56:14 & F845M & 15502 & Simon \\
2019-09-28 & 17:05:23 & F845M & 15502 & Simon \\
2019-09-28 & 21:42:19 & F845M & 15502 & Simon \\
2019-09-28 & 21:51:28 & F845M & 15502 & Simon \\
2019-09-29 & 00:53:02 & F845M & 15502 & Simon \\
2019-09-29 & 01:02:11 & F845M & 15502 & Simon \\
2019-09-29 & 04:03:46 & F845M & 15502 & Simon \\
2019-09-29 & 04:12:55 & F845M & 15502 & Simon \\
2019-09-29 & 07:14:28 & F845M & 15502 & Simon \\
2019-09-29 & 07:23:37 & F845M & 15502 & Simon \\
2019-09-29 & 12:00:33 & F845M & 15502 & Simon \\
2019-09-29 & 12:00:33 & F845M & 15502 & Simon \\
2020-01-07 & 19:16:42 & F845M & 16057 & Wong \\
2020-01-07 & 19:24:50 & F845M & 16057 & Wong \\
2020-01-07 & 19:49:11 & F845M & 16057 & Wong \\
2020-01-07 & 19:53:13 & F845M & 16057 & Wong \\
2020-01-07 & 22:27:22 & F845M & 16057 & Wong \\
2020-01-07 & 22:35:30 & F845M & 16057 & Wong \\
2020-01-07 & 22:59:51 & F845M & 16057 & Wong \\
2020-01-07 & 23:03:53 & F845M & 16057 & Wong \\
2020-01-08 & 01:38:02 & F845M & 16057 & Wong \\
2020-01-08 & 01:46:10 & F845M & 16057 & Wong \\
2020-01-08 & 02:10:31 & F845M & 16057 & Wong \\
2020-01-08 & 02:14:33 & F845M & 16057 & Wong \\
2020-01-08 & 04:48:42 & F845M & 16057 & Wong \\
2020-01-08 & 04:56:50 & F845M & 16057 & Wong \\
2020-01-08 & 05:21:11 & F845M & 16057 & Wong \\
2020-01-08 & 05:25:13 & F845M & 16057 & Wong \\
2020-01-08 & 07:59:20 & F845M & 16057 & Wong \\
2020-01-08 & 08:07:28 & F845M & 16057 & Wong \\
2020-01-08 & 08:31:49 & F845M & 16057 & Wong \\
2020-01-08 & 08:35:51 & F845M & 16057 & Wong \\
2020-06-23 & 05:29:47 & F845M & 16084 & Wong \\
2020-06-23 & 05:37:55 & F845M & 16084 & Wong \\
2020-06-23 & 06:02:16 & F845M & 16084 & Wong \\
2020-06-23 & 06:06:18 & F845M & 16084 & Wong \\
2020-06-23 & 08:40:28 & F845M & 16084 & Wong \\
2020-06-23 & 08:48:36 & F845M & 16084 & Wong \\
2020-06-23 & 09:12:57 & F845M & 16084 & Wong \\
2020-06-23 & 09:16:59 & F845M & 16084 & Wong \\
2020-06-23 & 11:51:08 & F845M & 16084 & Wong \\
2020-06-23 & 11:59:16 & F845M & 16084 & Wong \\
2020-06-23 & 12:23:37 & F845M & 16084 & Wong \\
2020-06-23 & 12:27:39 & F845M & 16084 & Wong \\
2020-06-23 & 15:01:50 & F845M & 16084 & Wong \\
2020-06-23 & 15:09:58 & F845M & 16084 & Wong \\
2020-06-23 & 15:34:19 & F845M & 16084 & Wong \\
2020-06-23 & 15:38:21 & F845M & 16084 & Wong \\
2020-06-23 & 18:12:30 & F845M & 16084 & Wong \\
2020-06-23 & 18:20:38 & F845M & 16084 & Wong \\
2020-06-23 & 18:44:59 & F845M & 16084 & Wong \\
2020-06-23 & 18:49:01 & F845M & 16084 & Wong \\
2020-12-12 & 07:02:34 & F845M & 16454 & Wong \\
2020-12-12 & 07:06:36 & F845M & 16454 & Wong \\
2020-12-12 & 07:12:10 & F845M & 16454 & Wong \\
2020-12-12 & 07:17:44 & F845M & 16454 & Wong \\
2020-12-12 & 13:23:54 & F845M & 16454 & Wong \\
2020-12-12 & 13:27:56 & F845M & 16454 & Wong \\
2020-12-12 & 13:33:30 & F845M & 16454 & Wong \\
2020-12-12 & 13:39:04 & F845M & 16454 & Wong \\
2020-12-12 & 16:34:34 & F845M & 16454 & Wong \\
2020-12-12 & 16:38:36 & F845M & 16454 & Wong \\
2020-12-12 & 16:44:10 & F845M & 16454 & Wong \\
2020-12-12 & 16:49:44 & F845M & 16454 & Wong \\
2020-12-12 & 19:45:13 & F845M & 16454 & Wong \\
2020-12-12 & 19:49:15 & F845M & 16454 & Wong \\
2020-12-12 & 19:54:49 & F845M & 16454 & Wong \\
2020-12-12 & 20:00:23 & F845M & 16454 & Wong \\
2020-12-13 & 08:27:53 & F845M & 16454 & Wong \\
2020-12-13 & 08:31:55 & F845M & 16454 & Wong \\
2020-12-13 & 08:37:29 & F845M & 16454 & Wong \\
2020-12-13 & 08:43:03 & F845M & 16454 & Wong \\
2020-12-13 & 11:38:31 & F845M & 16454 & Wong \\
2020-12-13 & 11:42:33 & F845M & 16454 & Wong \\
2020-12-13 & 11:48:07 & F845M & 16454 & Wong \\
\enddata
\end{deluxetable}

\startlongtable
\begin{deluxetable}{cccccc}
\tablecaption{Keck Data Used in this Paper\label{tab:keck_lick_data_table}}
\tablewidth{0pt}
\tablehead{ 
\colhead{Date} & \colhead{Time} & \colhead{Filter} & \colhead{Calibrated in Section \ref{sec:bootstrap}?} & \colhead{PI/Observers} & \colhead{Twilight?}
}
\startdata
2002-08-02 & 09:29:50 & H  & Y & Gibbard \\
2002-08-02 & 09:57:42 & K' & Y & Gibbard \\
2002-08-03 & 09:47:46 & H  & Y & Gibbard \\
2002-08-03 & 10:00:25 & K' & Y & Gibbard \\
2002-08-04 & 11:22:27 & H  & Y & Hammel \\
2002-08-04 & 11:27:17 & K' & Y & Hammel \\
2002-08-05 & 09:00:58 & K' & Y & Hammel \\
2002-08-05 & 11:22:44 & H  & Y & Hammel \\
2003-08-15 & 08:02:33 & H  & Y & Sromovsky \\
2003-08-15 & 08:13:03 & K' & Y & Sromovsky \\
2003-08-15 & 11:54:56 & H  & Y & Sromovsky \\
2003-08-15 & 12:05:09 & K' & Y & Sromovsky \\
2003-08-16 & 09:16:11 & H  & Y & Sromovsky \\
2003-08-16 & 09:34:01 & K' & Y & Sromovsky \\
2003-08-16 & 11:27:41 & H  & Y & Sromovsky \\
2003-08-16 & 11:44:09 & K' & Y & Sromovsky \\
2003-10-03 & 06:33:13 & H  &   & Hammel \\
2003-10-03 & 07:09:45 & K' &   & Hammel \\
2003-10-04 & 05:47:22 & H  &   & Hammel \\
2003-10-04 & 06:10:10 & K' &   & Hammel \\
2003-10-04 & 08:04:22 & H  &   & Hammel \\
2003-10-04 & 08:30:28 & K' &   & Hammel \\
2003-10-05 & 06:00:35 & K' &   & Gibbard, de Pater \\
2003-10-05 & 06:12:13 & H  &   & Gibbard, de Pater \\
2003-10-05 & 08:06:03 & H  &   & Gibbard, de Pater \\
2003-10-05 & 08:30:37 & K' &   & Gibbard, de Pater \\
2003-10-06 & 06:25:04 & H  &   & Gibbard, de Pater \\
2003-10-06 & 07:45:18 & K' &   & Gibbard, de Pater \\
2003-10-06 & 07:59:38 & H  &   & Gibbard, de Pater \\
2004-07-03 & 14:10:19 & H  &   & Hammel, de Pater \\
2004-07-04 & 10:36:42 & H  &   & Hammel, de Pater \\
2004-07-08 & 14:18:12 & H  &   & de Pater, Hammel \\
2004-07-09 & 11:51:38 & H  &   & de Pater, Hammel \\
2004-07-11 & 11:49:59 & H  &   & Sromovsky, Fry \\
2004-07-11 & 11:54:25 & K' & Y & Sromovsky, Fry \\
2004-07-11 & 14:25:57 & H  &   & Sromovsky, Fry \\
2004-07-11 & 14:30:26 & K' & Y & Sromovsky, Fry \\
2004-07-12 & 11:28:20 & H  & Y & Sromovsky, Fry \\
2004-07-12 & 11:33:14 & K' & Y & Sromovsky, Fry \\
2004-07-12 & 14:16:25 & H  & Y & Sromovsky, Fry \\
2004-07-12 & 14:21:08 & K' & Y & Sromovsky, Fry \\
2004-08-11 & 06:01:06 & H  & Y & Sromovsky, Fry \\
2004-08-11 & 06:06:20 & K' & Y & Sromovsky, Fry \\
2004-08-11 & 11:09:59 & H  & Y & Sromovsky, Fry \\
2004-08-11 & 11:14:52 & K' & Y & Sromovsky, Fry \\
2004-08-12 & 06:01:31 & H  & Y & Sromovsky, Fry \\
2004-08-12 & 06:12:17 & K' & Y & Sromovsky, Fry \\
2004-08-12 & 12:29:47 & K' & Y & Sromovsky, Fry \\
2005-07-05 & 11:00:07 & K' & Y & Hammel, de Pater \\
2005-07-05 & 11:17:15 & H  & Y & Hammel, de Pater \\
2005-07-06 & 13:18:51 & K' & Y & Hammel, de Pater \\
2005-07-06 & 13:26:25 & H  & Y & Hammel, de Pater \\
2005-08-14 & 11:55:23 & H  & Y & Sromovsky, Fry \\
2005-08-14 & 12:00:12 & K' & Y & Sromovsky, Fry \\
2005-08-15 & 11:12:19 & H  & Y & Sromovsky, Fry \\
2005-08-15 & 11:17:41 & K' & Y & Sromovsky, Fry \\
2005-08-23 & 06:52:52 & H  &  & de Pater \\
2005-08-23 & 07:06:48 & K' &  & de Pater \\
2005-10-29 & 06:17:32 & K' & Y & Hammel, de Pater \\
2005-10-29 & 06:22:47 & H  & Y & Hammel, de Pater \\
2006-07-29 & 11:07:48 & H  & Y & Sromovsky,Fry \\
2006-07-30 & 11:14:52 & H  & Y & Sromovsky,Fry \\
2006-07-30 & 11:19:25 & K' & Y & Sromovsky,Fry \\
2007-05-28 & 12:38:26 & K' & Y  & de Pater\\
2007-05-28 & 13:15:16 & H  & Y  & de Pater\\
2007-07-26 & 11:44:48 & K' &   & de Pater\\
2007-07-26 & 11:50:45 & H  &   & de Pater\\
2007-07-27 & 10:36:44 & K' &   & de Pater\\
2007-07-27 & 10:48:12 & H  &   & de Pater\\
2007-07-28 & 01:40:27 & H  &   & de Pater\\
2007-07-28 & 11:22:31 & K' &   & de Pater\\
2007-08-09 & 10:33:08 & K' &   & de Pater, Gibbard, Showalter\\
2007-08-09 & 10:38:15 & H  &   & de Pater, Gibbard, Showalter\\
2007-08-14 & 07:09:20 & H  &   & de Pater, Gibbard, Showalter\\
2007-08-14 & 07:18:27 & K' &   & de Pater, Gibbard, Showalter\\
2008-10-05 & 05:05:55 & H  &   & de Pater \\
2008-10-05 & 05:11:25 & K' &   & de Pater \\
2008-10-06 & 06:10:14 & H  &   & de Pater \\
2008-10-06 & 09:10:25 & H  &   & de Pater \\
2008-10-09 & 06:07:16 & H  &   & Hammel \\
2008-10-09 & 06:18:21 & K' &   & Hammel \\
2009-07-14 & 11:09:04 & K' &   & de Pater \\
2009-07-14 & 11:22:20 & H  &   & de Pater \\
2009-07-16 & 10:43:23 & H  &   & Hammel \\
2009-07-16 & 10:56:39 & K' &   & Hammel \\
2009-07-25 & 10:01:45 & K' & Y & de Pater\\
2009-07-25 & 10:12:01 & H  & Y & de Pater\\
2009-07-26 & 10:50:06 & H  & Y & Hammel \\
2009-07-26 & 10:55:35 & K' & Y & Hammel \\
2010-07-27 & 10:40:09 & H  & Y & de Pater \\
2010-07-27 & 10:51:04 & K' & Y & de Pater \\
2010-07-28 & 10:38:12 & H  & Y & de Pater \\
2010-07-28 & 10:42:16 & K' & Y & de Pater \\
2010-08-21 & 10:38:24 & K' &   & de Pater \\
2010-08-21 & 10:43:44 & H  &   & de Pater \\
2010-08-22 & 11:06:06 & H  &   & de Pater \\
2010-08-22 & 11:11:32 & K' &   & de Pater \\
2011-07-26 & 10:36:26 & H  & Y & Sromovsky, Fry \\
2011-07-26 & 10:43:06 & K' & Y & Sromovsky, Fry \\
2011-07-28 & 10:34:50 & H  & Y & de Pater \\
2011-07-28 & 10:39:58 & K' & Y & de Pater \\
2012-06-26 & 13:21:12 & H  & Y & Engineering \\
2012-06-26 & 13:37:44 & K' & Y & Engineering \\
2012-07-14 & 10:36:03 & K' & Y & de Pater \\
2012-07-14 & 10:55:07 & H  & Y & de Pater \\
2012-07-25 & 09:13:35 & H  & Y & Sromovsky, Fry \\
2012-07-25 & 09:33:14 & K' & Y & Sromovsky, Fry \\
2012-07-30 & 10:49:14 & K' & Y & de Pater \\
2012-07-30 & 11:03:12 & H  & Y & de Pater \\
2013-07-03 & 10:50:11 & H  &   & de Pater \\
2013-07-03 & 10:59:55 & K' &   & de Pater \\
2013-07-03 & 12:50:18 & K' &   & de Pater \\
2013-07-03 & 12:58:46 & H  &   & de Pater \\
2013-07-03 & 15:03:11 & H  &   & de Pater \\
2013-07-31 & 13:17:44 & K' & Y & de Pater \\
2013-07-31 & 13:25:11 & H  & Y & de Pater \\
2014-08-05 & 10:53:27 & H  &   & de Pater \\
2014-08-05 & 11:00:11 & K' &   & de Pater \\
2014-08-06 & 10:46:42 & H  &   & de Pater \\
2014-08-06 & 10:51:57 & K' &   & de Pater \\
2014-08-20 & 08:15:40 & H  &   & de Pater \\
2014-08-20 & 11:32:39 & H  &   & de Pater \\
2014-08-20 & 13:21:43 & H  &   & de Pater \\
2015-08-29 & 12:30:09 & H  &   & de Pater \\
2015-08-29 & 12:35:15 & K' &   & de Pater \\
2015-08-30 & 12:08:30 & H  &   & de Pater \\
2015-08-30 & 12:13:37 & K' &   & de Pater \\
2015-12-25 & 04:25:06 & H  &   & de Pater \\
2015-12-25 & 04:29:55 & K' &   & de Pater \\
2015-12-25 & 05:38:23 & H  &   & de Pater \\
2016-09-12 & 10:14:07 & H  & Y & Baranec\\
2016-10-16 & 07:19:32 & H  & Y & Baranec\\
2017-06-26 & 14:52:33 & H  &   & de Kleer, Molter \\
2017-06-26 & 14:57:51 & K' &   & de Kleer, Molter\\
2017-07-02 & 12:06:37 & H  &   & Team Keck \\
2017-07-02 & 12:17:35 & K' &   & Team Keck \\
2017-07-16 & 15:08:03 & H  & Y & Puniwai \\
2017-07-24 & 12:53:49 & H  & Y & de Pater, Tollefson \\
2017-07-24 & 13:01:04 & K' & Y & de Pater, Tollefson \\
2017-07-25 & 15:14:44 & H  &   & de Pater, Molter \\
2017-07-25 & 15:20:22 & K' &   & de Pater, Molter \\
2017-08-03 & 13:25:53 & H  & Y & Molter, Jordan & Y \\
2017-08-03 & 15:26:23 & H  & Y & Molter, Jordan & Y \\
2017-08-10 & 12:45:42 & K' & Y & Baranec, Salama & Y \\
2017-08-25 & 11:27:16 & H  &   & de Pater, Molter, Tollefson \\
2017-08-25 & 11:30:46 & K' &   & de Pater, Molter, Tollefson\\
2017-08-26 & 10:33:10 & H  &   & de Pater, Molter, Tollefson \\
2017-08-26 & 10:38:49 & K' &   & de Pater, Molter, Tollefson \\
2017-08-26 & 13:36:10 & H  & Y & de Pater, Molter, Tollefson \\
2017-09-03 & 10:31:36 & H  & Y & de Pater, Molter, Tollefson \\
2017-09-03 & 10:37:02 & K' & Y & de Pater, Molter, Tollefson \\
2017-09-03 & 12:57:19 & H  & Y & de Pater, Molter, Tollefson \\
2017-09-04 & 10:32:34 & H  & Y & Team Keck & Y \\
2017-09-04 & 10:37:35 & K' & Y & Team Keck & Y \\
2017-09-04 & 12:40:40 & H  & Y & Team Keck & Y \\
2017-09-27 & 04:56:27 & H  & Y & Mcilroy, Magnier & Y \\
2017-09-27 & 05:02:01 & K' & Y & Mcilroy, Magnier & Y \\
2017-10-06 & 10:48:40 & K' & Y & Aycock, Ragland & Y \\
2017-10-06 & 10:54:32 & H  & Y & Aycock, Ragland & Y \\
2017-11-08 & 04:14:10 & H  & Y & Alvarez, Licandro & Y \\
2017-11-08 & 04:19:22 & K' & Y & Alvarez, Licandro & Y \\
2018-01-10 & 04:38:21 & H  & Y & Puniwai, McPartland & Y \\
2018-01-10 & 04:43:26 & K' & Y & Puniwai, McPartland & Y \\
2018-05-22 & 14:57:04 & H  & Y & Alvarez, Mcilroy, Bennett & Y \\
2018-05-22 & 15:02:14 & K' & Y & Alvarez, Mcilroy, Bennett & Y \\
2018-05-23 & 14:55:28 & H  & Y & Mcilroy, Ridenour, Alvarez, Bennett & Y \\
2018-05-23 & 15:00:38 & K' & Y & Mcilroy, Ridenour, Alvarez, Bennett & Y \\
2018-05-25 & 15:00:03 & H  & Y & Aycock, Ridenour, Bennett & Y \\
2018-05-25 & 15:05:33 & K' & Y & Aycock, Ridenour, Bennett & Y \\
2018-05-26 & 14:46:03 & H  & Y & Aycock, Ridenour, Bennett & Y \\
2018-05-26 & 14:51:13 & K' & Y & Aycock, Ridenour, Bennett & Y \\
2018-05-27 & 14:49:56 & H  & Y & Aycock, Bennett & Y \\
2018-05-27 & 14:55:21 & K' & Y & Aycock, Bennett & Y \\
2018-06-11 & 14:37:52 & H  & Y & Stickel, Alvarez, Hu & Y \\
2018-06-11 & 14:43:06 & K' & Y & Stickel, Alvarez, Hu & Y \\
2019-06-15 & 14:28:43 & H  & Y & de Kleer, Pelletier \\
2019-06-15 & 14:33:45 & K' & Y & de Kleer, Pelletier\\
2019-06-16 & 14:56:27 & H  & Y & Pelletier & Y \\
2019-06-16 & 15:01:48 & K' & Y & Pelletier & Y \\
2019-06-17 & 14:54:24 & H  & Y & Pelletier, Gaidos & Y \\
2019-06-17 & 14:59:51 & K' & Y & Pelletier, Gaidos & Y \\
2019-07-04 & 15:09:31 & H  & Y & Team Keck & Y \\
2019-07-04 & 15:14:38 & K' & Y & Team Keck & Y \\
2019-09-10 & 11:25:14 & H  & Y & Sromovsky, Fry, de Pater\\
2019-09-10 & 11:30:44 & K' & Y & Sromovsky, Fry, de Pater\\
2019-09-11 & 11:19:00 & H  & Y & Sromovsky, Fry, de Pater\\
2019-09-11 & 11:24:20 & K' & Y & Sromovsky, Fry, de Pater\\
2019-10-28 & 08:28:59 & H  &   & de Pater\\
2019-10-28 & 08:33:55 & K' &   & de Pater\\
2019-11-04 & 04:15:52 & H  &  & de Pater\\
2019-11-04 & 04:21:04 & K' &  & de Pater\\
2019-11-04 & 04:22:26 & K' &  & de Pater\\
2020-05-19 & 15:13:01 & H  & Y & Hershley, Steidel, Chen & Y\\
2020-05-19 & 15:18:03 & K' & Y & Hershley, Steidel, Chen & Y\\
2020-08-11 & 15:28:00 & H  & Y & Renaud-Kim, Stockton & Y\\
2020-08-11 & 15:33:12 & K' & Y & Renaud-Kim, Stockton & Y\\
2020-08-15 & 15:13:47 & H  & Y & Wilburn, Alvarez, Cowie, Barger & Y \\
2020-08-15 & 15:18:57 & K' & Y & Wilburn, Alvarez, Cowie, Barger & Y \\
2020-09-08 & 07:13:46 & H  & Y & de Kleer, Camarca \\
2020-09-08 & 07:18:56 & K' & Y & de Kleer, Camarca \\
2020-11-06 & 05:49:12 & H  & Y & Alvarez, Pelletier & Y \\
2020-11-06 & 05:55:34 & H  & Y & Alvarez, Pelletier & Y \\
2020-11-06 & 06:00:52 & K' & Y & Alvarez, Pelletier & Y \\
2020-11-06 & 06:33:01 & H  & Y & Alvarez, Pelletier & Y \\
2020-11-06 & 06:38:14 & K' & Y & Alvarez, Pelletier & Y \\
2020-11-24 & 05:08:13 & H  & Y & de Pater \\
2020-11-24 & 06:41:23 & K' & Y & de Pater \\
2020-11-24 & 06:49:21 & H  & Y & de Pater \\
2021-05-11 & 15:14:44 & H  & Y & Rostopchina, Prochaska & Y\\
2021-05-11 & 15:19:59 & K' & Y & Rostopchina, Prochaska & Y \\
2021-07-21 & 14:32:45 & H  & Y & de Pater \\
2021-07-21 & 14:37:25 & K' & Y & de Pater \\
2021-07-21 & 14:45:34 & K' & Y & de Pater\\
2021-09-23 & 10:59:11 & H  &   & Sromovsky, Fry, de Pater \\
2021-09-23 & 11:03:42 & K' &   & Sromovsky, Fry, de Pater \\
2021-10-07 & 04:58:32 & H  &   & de Pater, Molter \\
2021-10-07 & 05:03:26 & K' &   & de Pater, Molter \\
2021-10-07 & 08:40:25 & H  &   & de Pater, Molter \\
2021-10-07 & 08:47:56 & H  &   & de Pater, Molter \\
2021-10-07 & 08:54:48 & H  &   & de Pater, Molter \\
2021-10-07 & 09:02:03 & H  &   & de Pater, Molter \\
2021-10-07 & 09:08:58 & H  &   & de Pater, Molter \\
2021-10-07 & 09:16:02 & H  &   & de Pater, Molter\\
2021-10-08 & 04:31:49 & H  &   & de Pater, Molter\\
2021-10-08 & 04:37:10 & K' &   & de Pater, Molter\\
2021-10-08 & 05:11:41 & H  &   & de Pater, Molter\\
2021-10-08 & 05:25:41 & H  &   & de Pater, Molter\\
2021-10-08 & 05:40:12 & K'  &   & de Pater, Molter\\
2021-10-08 & 05:54:19 & K'  &   & de Pater, Molter\\
2021-10-08 & 06:08:26 & K'  &   & de Pater, Molter\\
2021-10-08 & 06:22:44 & K'  &   & de Pater, Molter\\
2021-10-08 & 06:29:48 & K'  &   & de Pater, Molter\\
2021-10-08 & 06:37:10 & H  &   & de Pater, Molter\\
2021-10-08 & 07:05:11 & H  &   & de Pater, Molter\\
2021-10-08 & 07:33:01 & H  &   & de Pater, Molter\\
2021-10-08 & 07:54:09 & H  &   & de Pater, Molter\\
2021-10-08 & 08:42:57 & H  &   & de Pater, Molter\\
2021-10-08 & 09:04:34 & H  &   & de Pater, Molter\\
2021-10-08 & 09:18:36 & H  &   & de Pater, Molter\\
2022-05-17 & 14:50:34 & H  & Y & Alvarez, Aycock &  Y \\
2022-05-17 & 14:55:13 & K'  & Y & Alvarez, Aycock &  Y \\
2022-06-19 & 14:42:13 & H  & Y & Renaud-Kim, Alvarez & Y \\
2022-06-19 & 14:47:09 & K'  & Y & Renaud-Kim, Alvarez & Y \\
2022-06-29 & 14:44:01 & H  & Y & Aycock, Taylor & Y \\
2022-06-29 & 14:48:41 & K'  & Y & Aycock, Taylor & Y \\
2022-07-25 & 15:18:46 & H  & Y & de Pater \\
2022-08-03 & 15:28:36 & H  & Y & de Pater \\
2022-08-15 & 08:46:14 & H  &   & de Pater \\
2022-08-15 & 08:51:05 & K' &   & de Pater \\
2022-08-15 & 14:05:52 & H  &   & de Pater \\
2022-08-15 & 14:10:39 & K' &   & de Pater \\
2022-08-15 & 15:32:43 & H  &   & de Pater \\
2022-08-16 & 08:40:56 & H  &   & de Pater \\
2022-08-16 & 08:45:35 & K'  &   & de Pater \\
2022-08-16 & 14:33:02 & H  &   & de Pater \\
2022-08-16 & 14:37:07 & K'  &   & de Pater \\
2022-09-06 & 10:57:22 & H  &   & de Pater, Fry, Sromovsky \\
2022-09-06 & 11:02:15 & K'  &   & de Pater, Fry, Sromovsky \\
2022-09-06 & 13:32:57 & H  &   & de Pater, Fry, Sromovsky \\
2022-09-06 & 13:37:43 & K'  &   & de Pater, Fry, Sromovsky \\
2022-09-07 & 10:59:30 & H  &   & de Pater, Fry, Sromovsky \\
2022-09-07 & 11:04:26 & K'  &   & de Pater, Fry, Sromovsky \\
2022-09-07 & 13:00:13 & H  &   & de Pater, Fry, Sromovsky \\
2022-09-07 & 13:05:38 & K'  &   & de Pater, Fry, Sromovsky \\
2022-09-11 & 11:04:32 & K'  &  Y & Sromovsky, Fry, de Pater \\
2022-09-12 & 12:06:15 & H  &   & Sromovsky, Fry, de Pater \\
2022-09-12 & 12:28:38 & K'  &   & Sromovsky, Fry, de Pater \\
\enddata
\end{deluxetable}

\startlongtable
\begin{deluxetable}{cccc}
\tablecaption{Lick Data Used in this Paper\label{tab:keck_lick_data_table}}
\tablewidth{0pt}
\tablehead{ 
\colhead{Date} & \colhead{Time} & \colhead{Filter} & \colhead{PI/Observers}
}
\startdata
 2018-09-27 & 04:25:27 & H & Gates, Rich \\
 2018-10-26 & 03:05:39 & H & Gates, Rich \\
 2018-11-25 & 03:29:13 & H & Gates, Rich\\
 2019-07-22 & 11:29:37 & H & Gates, Ammons \\
 2019-07-23 & 11:55:42 & H & Lynam, Rich \\
 2019-08-17 & 11:46:00 & H & Gates, Gonzales \\
 2019-08-20 & 11:31:42 & H & Lynam, Rich \\
 2019-08-21 & 09:43:27 & H & Lynam, Theissen \\
 2019-09-14 & 08:45:57 & H & Gates, Giacalone, Dressing \\
 2019-09-15 & 08:34:32 & H & Gates, Giacalone \\
\enddata
\end{deluxetable}

\bibliography{references}{}
\bibliographystyle{aasjournal}

\end{document}